\documentclass[12pt]{article}

\usepackage{graphicx}
\usepackage{amsfonts}
\usepackage{amsmath}
\usepackage{amssymb}
\usepackage{cite}

\numberwithin{equation}{section}

\def\spa#1{\phantom{\fbox{\rule[-#1cm]{0cm}{0cm}}}}

\def\eq#1{(\ref{#1})}
\def\[#1\]{\begin{align}#1\end{align}}

\def\nn{\nonumber}
\def\CR{\nonumber \\}

\def\middlerightarrow{\rightarrow\hspace{-.18cm}-}

\def\rD{\overset{\rightarrow}{D}{}}
\def\lD{\overset{\leftarrow}{D}{}}

\def\hM{\hat M}
\def\hP{\hat P}

\def\hA{\hat A}
\def\hB{\hat B}
\def\hC{\hat C}

\def\hH{\hat H}
\def\hJ{\hat J}
\def\hD{\hat D}

\def\cH{{\cal H}}

\begin{document}

\hfuzz=100pt
\title{
\begin{flushright} 
\vspace{-1cm}
\small{YITP-15-75}
\end{flushright} 
{\Large 
\bf An $OSp$ extension of Canonical Tensor Model
}
}
\date{}
\author{Gaurav Narain$^a$\footnote{gaunarain@gmail.com} and 
Naoki Sasakura$^b$\footnote{sasakura@yukawa.kyoto-u.ac.jp} 
  \spa{0.5} \\
$^a${\small{\it Max-Planck Institute f{\"u}r Gravitational Physics (Albert-Einstein Institute), }}
\\ {\small{\it  Am M{\"u}hlenberg 1, D-14476 Potsdam-Golm, Germany   }}
\spa{.5}\\
$^{b}${\small{\it Yukawa Institute for Theoretical Physics,}}
\\ {\small{\it  Kyoto University, Kyoto 606-8502, Japan}}
\spa{.5}\\
}
\date{}

\maketitle
\centerline{}

\begin{abstract}
Tensor models are generalizations of matrix models, and are studied as discrete 
models of quantum gravity for arbitrary dimensions. Among them, the canonical 
tensor model (CTM for short) is a rank-three tensor model formulated as a totally 
constrained system with a number of first-class constraints, which have a similar 
algebraic structure as the constraints of the ADM formalism of general relativity. 
In this paper, we formulate a super-extension of CTM as an attempt to incorporate 
fermionic degrees of freedom. The kinematical symmetry group is extended from 
$O(N)$ to $OSp(N,\tilde N)$, and the constraints are constructed so that they 
form a first-class constraint super-Poisson algebra. This is a straightforward 
super-extension, and the constraints and their algebraic structure are formally 
unchanged from the purely bosonic case, except for the additional signs associated 
to the order of the fermionic indices and dynamical variables. 
However, this extension of CTM leads to the existence 
of negative norm states in the quantized case, and requires some future improvements 
as quantum gravity with fermions. On the other hand, since this is a straightforward super-extension,
various results obtained so far for the purely bosonic case are expected to have parallels also in the 
super-extended case, such as the exact physical wave functions and the connection to 
the dual statistical systems, i.e. randomly connected tensor networks.
\end{abstract}

\newpage

\renewcommand{\thefootnote}{\arabic{footnote}}
\setcounter{footnote}{0}

\section{Introduction}
\label{Introduction}
%
Tensor models \cite{Ambjorn:1990ge,Sasakura:1990fs,Godfrey:1990dt} 
were introduced with the hope to analytically describe simplicial quantum gravity 
for arbitrary dimensions\footnote{However, see \cite{Fukuma:2015haa,Fukuma:2015xja} 
as a recent approach to three-dimensional quantum gravity in terms of matrix models.} 
by extending the matrix models which successfully describe the two-dimensional 
simplicial quantum gravity\cite{DiFrancesco:1993nw}. Subsequently, tensor models 
with group-valued indices\cite{Boulatov:1992vp,Ooguri:1992eb}, called group field 
theories\cite{DePietri:1999bx,Freidel:2005qe,Oriti:2011jm} were introduced, 
and have extensively been studied especially in connection with loop quantum gravity.
Some serious problems of the original tensor models\cite{DePietri:2000ii, Sasakura:1990fs}
have been overcome by the advent of the colored tensor models\cite{Gurau:2009tw}, 
and various interesting concrete results on their properties have been obtained\cite{Gurau:2011xp}
(see for instance \cite{Benedetti:2015ara,Delepouve:2015nia,Bonzom:2015axa,Delepouve:2014hfa,
Nguyen:2014mga} for some recent developments). The colored tensor models have 
also stimulated the renormalization group analysis of the group field theories (see for instance 
\cite{Benedetti:2015yaa,Avohou:2015sia,Geloun:2015lta,Lahoche:2015ola,
Benedetti:2014qsa,Geloun:2014ema} for some recent developments).

One of the main themes of the study of tensor models or generally in 
quantum gravity is to pursue the physical mechanism for the generation of the 
classical space-time like our universe. In the tensor models above, spaces are 
represented by simplicial manifolds, which are generated as the duals to the 
Feynman diagrams in the perturbative treatments of tensor models. The large 
$N$ analyses of the colored tensor models have shown that the generated 
simplicial manifolds are dominated by branched polymers 
\cite{Bonzom:2011zz,Gurau:2013cbh,Gurau:2011xp}. Naively, the result 
would be an obstacle for the tensor models to become sensible models of 
our space-time, since branch polymers do not seem like an extended entity in large 
scales. In fact, there are some interesting active directions of efforts in the colored 
tensor models to overcome this difficulty in terms of dynamically realized symmetries 
\cite{Benedetti:2015ara,Delepouve:2015nia} and higher orders 
\cite{Bonzom:2015axa,Raasakka:2013eda,Dartois:2013sra,Kaminski:2013maa,Gurau:2013pca}.  
On the other hand, in Causal Dynamical Triangulation, which is a Lorentzian model of 
simplicial quantum gravity, it has been shown that de Sitter-like space-times,
similar to our actual universe, are generated\cite{Ambjorn:2004qm}. This success 
can be contrasted with the unsuccessful situation in Dynamical Triangulation, 
which is the original Euclidean model. The main difference of the two models 
is the existence of the causal time-like direction in the former. Therefore, this 
success would indicate the importance of a time-like direction in quantum gravity, 
and would raise the possibility of improving the tensor models above, which 
basically dealt with Euclidean cases, by incorporating a time-like direction.  

With the motivation above, one of the present authors has formulated a tensor 
model in the Hamiltonian formalism (canonical tensor model or CTM for short below) 
\cite{Sasakura:2011sq,Sasakura:2012fb,Sasakura:2013gxg}\footnote{See \cite{Oriti:2013aqa} 
for a Hamiltonian approach in the framework of group field theories.}. It has been 
formulated as a totally constrained system similar to the ADM formalism of general 
relativity \cite{Arnowitt:1960es,Arnowitt:1962hi,DeWitt:1967yk,Hojman:1976vp,Teitelboim:1987zz}. 
CTM has a close parallel with the ADM formalism in the sense that CTM has the 
analogue of Hamiltonian and Momentum constraints of the ADM. It is therefore 
expected to respect the central principle of general relativity, i.e. the space-time 
general covariance. Under some reasonable physical assumptions, the model with 
a canonical conjugate pair of rank-three symmetric real tensors, the minimal model, has been 
shown to be unique with a free real parameter \cite{Sasakura:2012fb}.

The analysis so far have revealed various fascinating properties of CTM.
The $N=1$ case of CTM was shown to be equivalent to the mini-superspace 
approximation of general relativity, and the free parameter mentioned above has 
been identified with the ``cosmological constant" \cite{Sasakura:2014gia}.
It has been shown that there exists a formal continuum limit in which 
the constraint Poisson algebra of CTM agrees with that of ADM \cite{Sasakura:2015pxa}.
In the analysis, the correspondence between the variables of CTM and the 
metric tensor field of general relativity has partially been obtained \cite{Sasakura:2015pxa}.
It has been found that CTM and statistical systems on random networks 
have intimate relations: the renormalization group flows of randomly connected 
tensor networks can be described by the Hamiltonian of CTM 
\cite{Sasakura:2015xxa,Sasakura:2014zwa,Sasakura:2014yoa}. The quantization 
of CTM is straightforward \cite{Sasakura:2013wza}, and one can obtain a 
number of exact physical wave functions by explicitly solving the set of 
partial differential equations representing the constraints for small $N$ 
\cite{Sasakura:2013wza,Narain:2014cya}, and also by exploiting the relation 
between CTM and randomly connected tensor networks
for general $N$ \cite{Narain:2014cya}. It has been observed 
that the physical wave functions have singular behaviors at the 
configurations of potential physical importance characterized by locality 
\cite{Sasakura:2013wza} or group symmetries \cite{Narain:2014cya}\footnote{Thus, 
in CTM, it would be plausible that physically important configurations are stressed 
by physical wave functions. This is in contrast with the situation in the Euclidean 
rank-three tensor models, in which the actions were initially fine-tuned so 
that the minimums generate physically sensible backgrounds 
\cite{Sasakura:2008pe,Sasakura:2009hs}.}. The above interesting 
properties of CTM obtained so far would ensure CTM be worth studying further.

The main purpose of the present paper is to present an attempt to incorporate 
fermionic degrees of freedom into CTM. The origin of spinning particles is a deep 
interesting question (see for instance \cite{Rempel:2015foa} as a recent publication 
and references therein), and it is not obvious what is the most promising way to 
incorporate such degrees of freedom in the case of CTM. In this paper, we 
make an attempt in a most straightforward manner to do this by making use
of Grassmann odd variables. We introduce odd-type indices, and assume 
that the tensorial variables with an odd number 
of odd-type indices have the Grassmann odd property. This provides a 
straightforward extension of CTM in the sense that the constraints and their algebra 
are formally similar to the purely bosonic case, except for the additional signs 
associated to the order of the odd indices and variables. Thus, the 
kinematical symmetry is extended from $O(N)$ to $OSp(N,\tilde N)$, and 
the Hamiltonian constraints form a first-class constraint algebra with the 
kinematical constraints for the super-extended symmetry. However, this simple 
extension suffers from negative norm states in the quantized case. Therefore, 
without some improvements in future, this super-extension cannot be considered 
as a valid model of quantum gravity with fermions. On the other hand, as 
in the purely bosonic case \cite{Sasakura:2015xxa,Sasakura:2014zwa,Sasakura:2014yoa},
this super-extended model can be connected to the inclusion of fermionic degrees of 
freedom into the dual statistical systems, randomly connected tensor 
networks, since the connection is based on its classical properties. We would 
also expect that the exact physical wave functions obtained previously 
\cite{Sasakura:2013wza,Narain:2014cya} can be super-extended.

The paper is organized as follows. 
In Section~\ref{sec:basic}, we write down the basic definitions concerning indices and variables 
with attentions on their Granssmann even or odd properties.
In Section~\ref{sec:graph}, we introduce some graphical representations which 
can describe the complications of the signs associated to ordering in simplified manners.
We also present some graphical identities which are useful for the computations in the subsequent sections.
In Section~\ref{sec:poisson}, we consider the super-extension of the constraints from the bosonic case, 
and explicitly compute the Poisson algebra among them to show that they form a first-class constraint
algebra. The constraints and their algebra are essentially the same as the purely bosonic case except for the signs.
In Section~\ref{sec:reality}, we impose the reality condition on the variables, and check the 
consistency.
In Section~\ref{sec:quantization}, we perform the quantization of the super-extended model. 
This is straightforward, and the quantized constraint algebra has basically the 
same form as the purely bosonic case. 
The final section is devoted to summary and discussions.
The details of the computations in Section~\ref{sec:poisson} are shown in the appendix.

\section{Basic definitions}
\label{sec:basic}
%
In this section, we introduce super-extensions of the dynamical variables 
and some related things. The notations used here are not exactly, but are largely 
based on the textbooks \cite{DeWitt:1992cy, Henneaux:1992ig}.

Let us consider that any index $a$ of a tensor runs through a set
of labels. The labels are classified by a $Z_2$ grade $|a|=0 \hbox{ or }1$,
where the total number of even ones with $|a|=0$ is given by $N$, while
that of odd ones with $|a|=1$ is by $\tilde N$.
Then, each component of a variable $A$ with $p$ indices 
is assumed to have a $Z_2$ grade defined by
\[
\left| A_{a_1a_2\ldots a_p}\right|=\sum_{i=1}^p |a_i| \hbox{ mod }2.
\]
The $Z_2$ grades of a variable has a very crucial role to play
in the tensor model in the sense that it determines the properties under 
the change of their orders, namely their even (bosonic) or 
odd (fermionic) properties, as
\[
A_{a_1\ldots a_p}  B_{b_1\ldots b_q} 
=(-1)^{|A_{a_1\ldots a_p}||B_{b_1\ldots b_q}|}B_{b_1\ldots b_q}A_{a_1\ldots a_p}   
&=(-1)^{\sum_{ij}|a_i||b_j|} B_{b_1\ldots b_q}A_{a_1\ldots a_p} \CR
&=(-1)^{\sum_{ij}a_i b_j}B_{b_1\ldots b_q}A_{a_1\ldots a_p},
\label{eq:orderfg}
\]
Here, in the second line, we have written an abbreviation frequently used in 
this paper. Obviously,
\[
(-1)^{2a}=1, \ (-1)^{aa}=(-1)^a
\]
hold for an arbitrary grade $a=0,1$. 

A totally symmetric three-index tensor, in the super-extended case, is defined by 
\[
M_{abc}=(-1)^{a(b+c)} M_{bca}=(-1)^{c(a+b)}M_{cab}=(-1)^{ba} M_{bac}
=(-1)^{bc} M_{acb}=(-1)^{ab+bc+ca}M_{cba},
\label{eq:defofsym}
\]
with the additional signs coming from the grade. This symmetric condition is 
imposed on both the dynamical variables $M,P$ of CTM. In the following, we 
will also be dealing with anti-symmetric matrices, which will appear as the 
coefficients for the kinematical constraints. In the purely bosonic case, 
the kinematical constraints are the generators of the orthogonal transformations, and the coefficients 
are anti-symmetric matrices. In the super-extended case, the anti-symmetric matrices 
are characterized by
\[
R_{ab}=-(-1)^{ab} R_{ba}
\label{eq:defofanti}
\]
with the additional sign. Thus, the kinematical symmetry is extended to an $OSp$ group.

At this point, we define a {\it constant bosonic symmetric} matrix 
and its inverse denoted by $\Omega$ satisfying
\[
&\Omega_{ab}=(-1)^{ab} \Omega_{ba}, \CR
&\Omega^{ab}=(-1)^{ab} \Omega^{ba}, \CR
&\Omega_{ab}\Omega^{bc}=\delta_a^c, \label{eq:omega}\\
&\Omega_{ab}=\Omega^{ab}=0, \hbox{ if }(-1)^{a+b}=-1. \nn
\]
Here, notice that $\Omega$ is invertible and non-singular, which also means 
that $\tilde N$ must be even. The last condition in \eq{eq:omega}, which 
requires $\Omega$ to be bosonic, may not be necessary in general, 
but we assume it for the simplicity of the index contractions defined below. 
$\Omega$ defines the relations between the variables with lower and 
upper indices as
\[
A^a=A_b\Omega^{ba}, \ A_a=A^b \Omega_{ba}, 
\]
and similarly for matrices and tensors. Thus, in the formalism developed 
here, $\Omega$ plays the role of a metric for index contractions.

The commutator of matrices will be defined in the usual manner and
is given by
\[
[A,B]_{ab}=A_a{}^{c}B_{cb}-B_a{}^c A_{cb}.
\label{eq:matrixcommutator}
\]

Following the standard procedure, the fundamental Poisson bracket is defined by
\[
&  \{M_{abc},P^{def} \} =- \{P_{abc},M^{def} \}=\delta_{abc}^{fed}, \label{eq:fundamentalpoisson}\\
& \{M_{abc},M^{def} \}=\{P_{abc},P^{def} \}=0,
\]
where 
\[
\delta_{abc}^{fed}&\equiv \delta_a^f\delta_b^e\delta_c^d+(-1)^{f(e+d)}\delta_a^e\delta_b^d\delta_c^f
+(-1)^{d(f+e)}\delta_a^d\delta_b^f\delta_c^e+(-1)^{ef}\delta_a^e\delta_b^f\delta_c^d  \CR
&\hspace{6cm}+(-1)^{de}\delta_a^f\delta_b^d\delta_c^e+(-1)^{de+ef+fd}\delta_a^d\delta_b^e\delta_c^f.
\label{eq:defofdelta3}
\] 
Here the signs are necessarily imposed in order to be consistent with the symmetric 
condition written in \eq{eq:defofsym}, and also note that the orders of the upper 
indices $def$ on $M,P$ are reversed on the rightmost expression in 
\eq{eq:fundamentalpoisson}. To define the Poisson bracket for general cases, 
let us start by introducing the following right and left first-order partial derivatives 
with respect to $M,P$ which satisfy 
\[
\rD^M_{abc} M^{def}=\rD^P_{abc} P^{def}=\delta_{abc}^{fed}, \CR
M_{abc}\lD_M^{def} =P_{abc}\lD_P^{def} =\delta_{abc}^{fed}.
\label{eq:defofpartialder}
\]
These are not simple partial derivatives due to the symmetric factors implicitly 
contained in \eq{eq:defofdelta3}\footnote{For instance, for the case $a=b=c=d=e=f=\hbox{even}$, 
\eq{eq:defofdelta3} produces a factor of 6, and \eq{eq:defofpartialder} differs by 
this factor from the usual partial derivatives. This factor necessarily appears for the compatibility between \eq{eq:defofsym}
and  the kinematical symmetry.}. One also has to be careful with 
the order of indices, which can introduce an additional sign in the computation.
The partial derivatives are assumed to satisfy the following Leibniz rule,
\[
&\rD^M_{abc} (A B)=(\rD^M_{abc} A) B+ (-1)^{|A|(a+b+c)} A( \rD^M_{abc} B),  \CR
&(A B)\lD_M^{abc} =A (B \lD_M^{abc})+ (-1)^{|B|(a+b+c)} ( A \lD_M^{abc})B,
\label{eq:Leibnizrule}
\]
and the same for $D^P,D_P$.
This is a sort of generalisation of the usual Leibniz rule taking into 
account the grade of the tensor indices. Then, using this, one can 
easily define the super-Poisson bracket for general cases by 
\[
\{ A, B\}=\frac{1}{6} \left[ (A \lD_M^{abc})(\rD^P_{cba} B)-(A \lD_P^{abc})(\rD^M_{cba} B)\right].
\] 
This indeed reproduces the fundamental Poisson bracket written in 
\eq{eq:fundamentalpoisson} by making use of the identity 
$\delta_{abc}^{ihg}\delta_{ihg}^{fed}=6 \delta_{abc}^{fed}$. 
This super-Poisson bracket has the following anti-commutative property,
\[
\{A,B\}=-(-1)^{|A||B|} \{ B,A \}.
\label{eq:commutationPoisson}
\]
From the Leibniz rule \eq{eq:Leibnizrule}, one can also derive the 
Leibniz rule for the Poisson bracket,
\[
&\{AB,C\}=A\{B,C\}+(-1)^{|B||C|} \{A,C\}B, \CR
&\{A,BC\}=\{A,B\}C+(-1)^{|A||B|} B\{A,C\}.
\label{eq:LeibnizPoisson}
\]
The Poisson bracket also satisfies the Jacobi identity,
\[
\{\{A,B\},C\}=\{A,\{B,C\}\}+(-1)^{BC} \{\{A,C\},B\}.
\]

\section{Graphical representation and identities}
\label{sec:graph}
%
The number of indices with grades involved in the tensorial computation is large.
This easily exponentiates the complexity in the manipulation and handling 
of the algebra, and therefore demands a methodology to do this task in a 
simplified manner. This will be achieved by making use of graphs 
to do index contractions and recognize the signs graphically.

\begin{figure}
\begin{center}
\includegraphics[scale=.7]{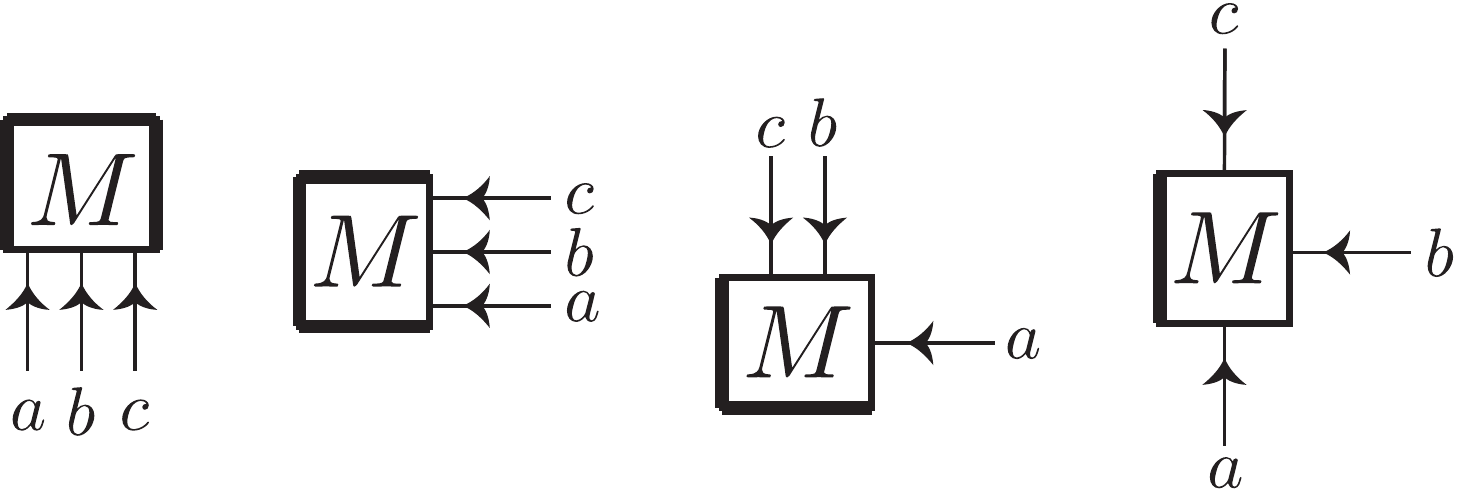}
\caption{Graphical representations of $M_{abc}$, which are supposed 
to be equivalent. Here vacant edges are drawn in thick bold lines to 
clarify the definition. Reading off the indices on a particular box is 
always done in a counter-clockwise manner. 
\label{fig:Mabc}}
\end{center}
\end{figure}

\begin{figure}
[ht]
\begin{center}
\includegraphics[scale=.6]{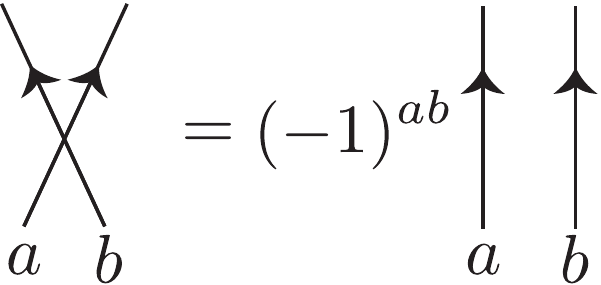}
\caption{The sign associated to a crossing.
\label{fig:cross}}
\end{center}
\end{figure}

\begin{figure}
[ht]
\begin{center}
\includegraphics[scale=.6]{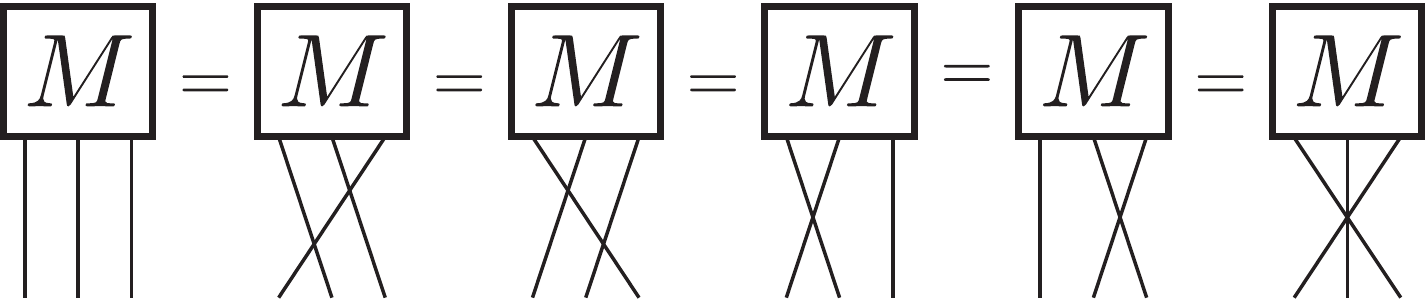}
\caption{The symmetric condition \eq{eq:defofsym} in terms of graphs.
This holds for $P$ as well.
\label{fig:symcond}}
\end{center}
\end{figure}

\begin{figure}
[ht]
\begin{center}
\includegraphics[scale=.6]{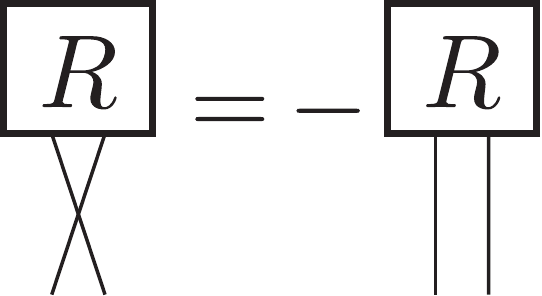}
\caption{The anti-symmetric condition \eq{eq:defofanti} in terms of graphs.
\label{fig:anti}}
\end{center}
\end{figure}

\begin{figure}
\begin{center}
\includegraphics[scale=1]{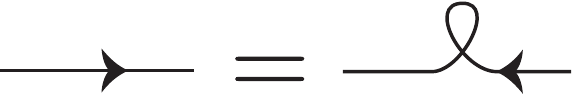}
\caption{The identity to reverse an orientation of an arrow corresponding to the last equation of 
\eq{eq:contractorder}. The sign is represented by the self-crossing due to $(-1)^{aa}=(-1)^a$.
\label{fig:contractorder}}
\end{center}
\end{figure}

The most basic expression can be written in the diagrammatic form as 
\[
A^aB_a:\ A\middlerightarrow\,B.
\label{eq:basicgraph}
\]
The generalization for multiple indices and variables can systematically be done.
Firstly, we introduce
\begin{itemize}
\item[(I)]
In a graph, variables are ordered in the horizontal direction, in the same order as appearing in a formula.  
\item[(II)]
A contracted pair of indices are connected by an arrow which starts from an upper index and ends 
on a lower one.
\end{itemize}
Here, the requirements stated in (I) and (II) respectively come 
from the dependence of overall signs on both the order of variables 
as in \eq{eq:orderfg} and the two ways of contractions as
\[
A^a B_a=(-1)^a B_a A^a=(-1)^aA_aB^a.
\label{eq:contractorder}
\]   
We introduce a box to represent a variable, rather than use the 
bare expression as in \eq{eq:basicgraph} for a vector.
For instance, a three-index tensor $M_{abc}$ is represented 
as in Figure~\ref{fig:Mabc}, where all the graphs are supposed to 
be equivalent.
In general, the correspondence between a multi-index variable 
and a graph is defined by the following rules:
\begin{itemize}
\item[(III)] A variable is represented by a box containing its label. 
\item[(IV)] Indices of a variable are assigned counter-clockwise on arrows attached to a box.
\item[(V)] The first index of a variable is assigned to the arrow which appears first after vacant edges of a box
in the counter clockwise direction.
Here, vacant edges mean those with no arrows attached, and must form a connected bunch for 
well-definedness (see Figure~\ref{fig:Mabc}).
\end{itemize}
It is convenient to associate a sign to a crossing of two lines as in Figure~\ref{fig:cross}.
A merit of this convention is to simplify the symmetric condition \eq{eq:defofsym}: 
signs coming from reordering of indices of a symmetric tensor 
can be represented by crossings as in Figure~\ref{fig:symcond}.
The convention also simplifies the anti-symmetric condition 
\eq{eq:defofanti} as in Figure~\ref{fig:anti}. We also have a graphical 
identity Figure~\ref{fig:contractorder} corresponding to the last equation 
of \eq{eq:contractorder}. 


\begin{figure}
[h]
\begin{center}
\includegraphics[scale=.5]{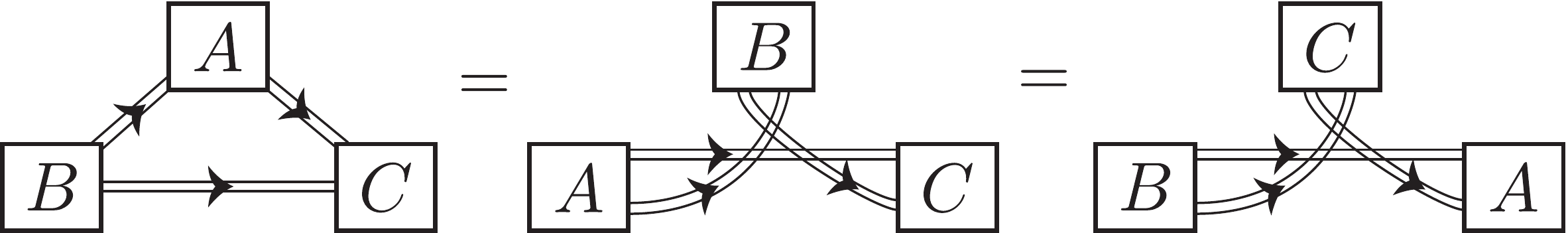}
\caption{An identity for reordering variables. In the above figure, 
$A,B,C$, and the double-line arrows represent bunches of variables and arrows, respectively.  
The identity still holds, even when some of $A,B,C$ or the double-line arrows are absent,
when bunches of arrows contain anti-parallel arrows, 
and when arrows cross in other ways.
The essential points are (i) Before and after reordering, connections and directions of arrows must be kept, 
except for those connecting reordered variables. 
(ii) As for the arrows connecting reordered variables, they must be reconnected with reversed 
directions in the ways shown in the figure. 
\label{fig:exchangerule}}
\end{center}
\end{figure}

\begin{figure}
\begin{center}
\includegraphics[scale=.5]{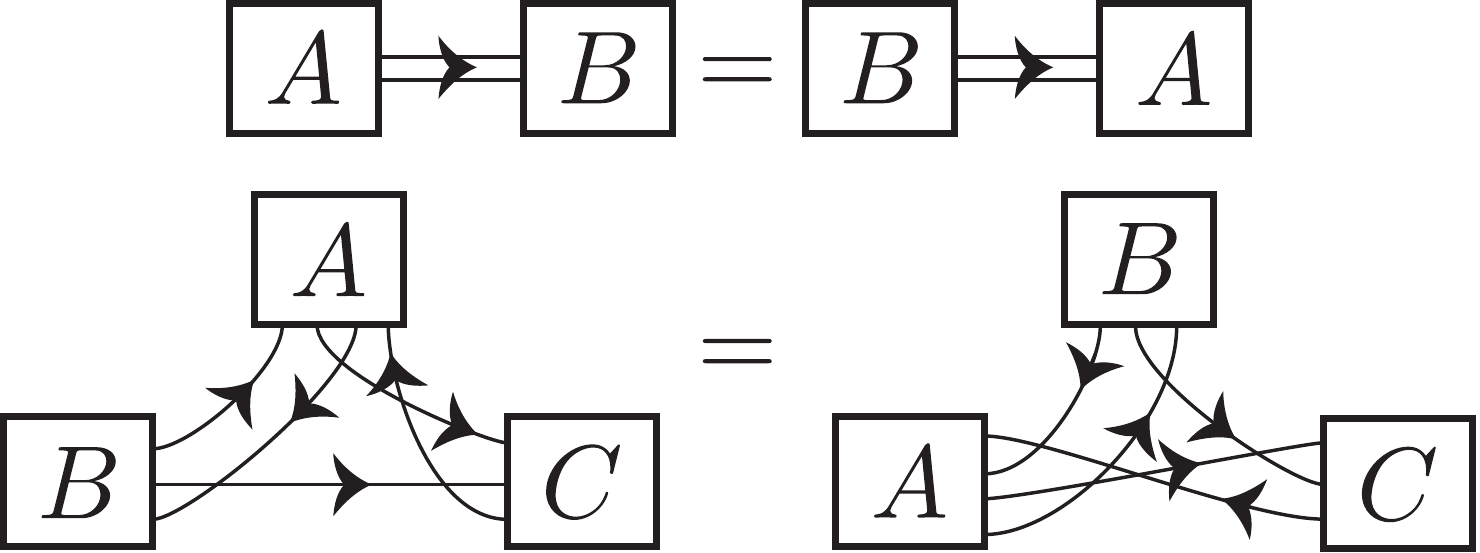}
\caption{Two examples of variants of the identity shown in Figure~\ref{fig:exchangerule}.
\label{fig:exchangeexample}}
\end{center}
\end{figure}

A very convenient graphical identity which will be used frequently in later computations is shown 
in Figure~\ref{fig:exchangerule}. This identity allows one to change order of variables. 
The identity for the simplest case of the first equality can be proven by
\[
A^{ab}B_{b}{}^{c}&=
(-1)^{(a+b)(b+c)} B_{b}{}^{c}A^{ab} \CR
&=(-1)^{(a+b)(b+c)+b} B^{bc}A^{a}{}_{b} \CR
&=(-1)^{ab+bc+ca} B^{bc}A^{a}{}_{b},
\label{eq:exchangerule}
\]
where we have used $|A^{ab}|=|a|+|b|,\ |B_b{}^c|=|b|+|c|\ \hbox{mod 2}$ for \eq{eq:orderfg} and 
\eq{eq:contractorder} for the index $b$.
Here, the nice thing is that the cumbersome sign in the last line of \eq{eq:exchangerule}
are accounted graphically by the crossings of the arrows.
The proof can obviously be generalized to other cases.
In fact, the identity shown in the Figure~\ref{fig:exchangerule} is a representative of various variants of the identity:
some of $A,B,C$ or the double-line arrows may be absent, double-line arrows may contain
anti-parallel arrows, and there may exist other crossings of arrows.    
As illustrations, two of such variants are shown in Figure~\ref{fig:exchangeexample}.
Basically, all the computations in the following sections can be performed 
by mostly using the variants of the identity and the basic properties of variables. 

\begin{figure}
\begin{center}
\includegraphics[scale=.8]{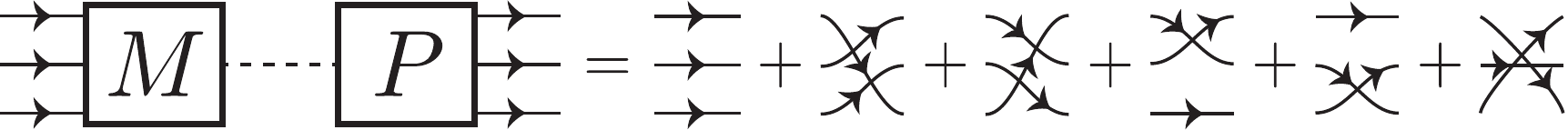}
\caption{The graphical representation of the fundamental Poisson bracket \eq{eq:fundamentalpoisson}. 
It is given by the summation of the six graphs in the figure.
The dashed line indicates the pair between which the fundamental Poisson bracket is taken.
The cumbersome signs in \eq{eq:fundamentalpoisson} are represented by the sign assignment 
Figure~\ref{fig:cross} to the crossings.
\label{fig:poisson}}
\end{center}
\end{figure}

Finally, we will show the graphical representation of the fundamental Poisson bracket.
By using the Leibniz rule \eq{eq:LeibnizPoisson}, 
the computation of a Poisson bracket between arbitrary quantities can be reduced to
summation of terms containing only the fundamental Poisson bracket.
The graphical representation of the fundamental Poisson bracket is given 
in Figure~\ref{fig:poisson}. As in the figure, we use a dashed line to indicate 
a pair of $M,P$ between which the fundamental Poisson bracket is taken. An advantage 
of this graphical representation is that the cumbersome signs appearing 
in \eq{eq:fundamentalpoisson} can be represented by the crossings due 
to the sign assignment Figure~\ref{fig:cross}.

\section{The Poisson algebra of the constraints}
\label{sec:poisson}

In this section, we define the constraints in the super-extended case, 
and show the Poisson algebra satisfied by them. It turns out that the constraints 
as well as the constraint Poisson algebra are formally unchanged from the purely bosonic 
case, in the sense that the overall structure of the constraints and the algebra satisfied 
by them is very much similar (modulo signs). This formal simplicity would be due 
to the validity and consistency of the basic properties 
imposed on the parameters and dynamical variables in the previous sections.   

By simply transferring the purely bosonic case to the present super-extended case
by taking care of index orderings, we consider the Hamiltonian constraints defined by
\[
{\cH}_a=\frac{1}{2} \left( P_a{}^{bc} P_c{}^{de}M_{edb} -\lambda M_a{}^b{}_b \right),
\]
where $\lambda$ is a real constant. It was shown in the purely bosonic case
that $\lambda$ can be interpreted as a cosmological constant, 
because the $N=1$ case in CTM exactly reproduces the mini-superspace 
approximation of general relativity with a cosmological constant
proportional to $\lambda$ \cite{Sasakura:2014gia}. For this reason, 
we call $\lambda$ in this case also to be the cosmological constant.

\begin{figure}
\begin{center}
\includegraphics[scale=.5]{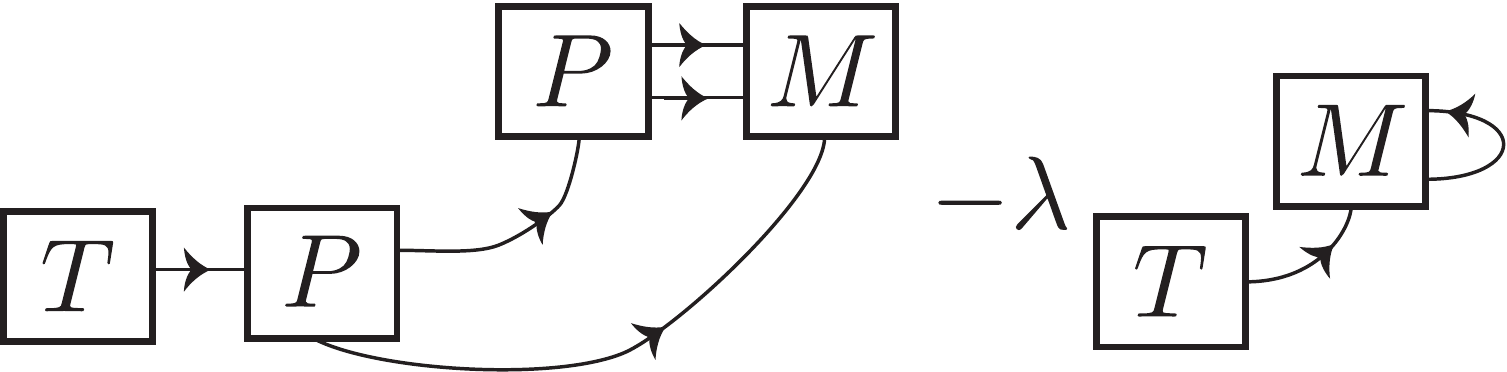}
\hfil
\includegraphics[scale=.5]{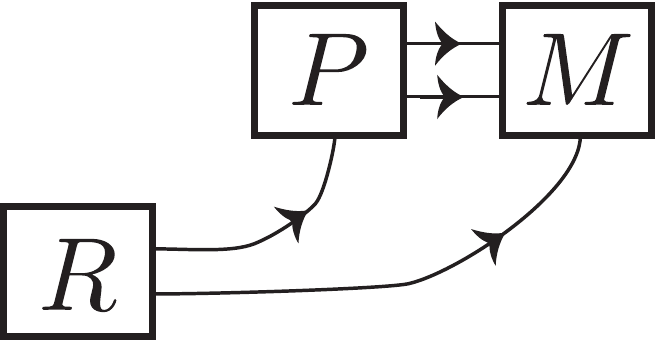}
\caption{The graphical representations of $H(T)$ in \eq{eq:hamiltonian} and
$J(R)$ in \eq{eq:momentum} are given by the left and the right figures, respectively, up to
the numerical factors.
\label{fig:hamiltonian}}
\end{center}
\end{figure}

It is convenient at this point to introduce a non-dynamical variable $T^a$, 
and consider
\[
H(T)=T^a {\cH}_a=\frac{1}{2} \left( T^a P_a{}^{bc} P_c{}^{de}M_{edb}
-\lambda T^a M_a{}^b{}_b\right).
\label{eq:hamiltonian}
\]
For convenience, we separately consider the two terms by defining $H_0(T)= \left. H(T) \right |_{\lambda=0}$.
Similarly, the momentum constraints are given by
\[
J(R)=-\frac{1}{2} R^{ab}P_{b}{}^{cd}M_{dca},
\label{eq:momentum}
\]
where $R$ is a non-dynamical matrix variable with the anti-symmetric 
property \eq{eq:defofanti}, and the overall minus sign is for convenience.
The graphical representations of the constraints are depicted diagrammatically 
in Figure~\ref{fig:hamiltonian} up to the numerical factors.

The constraint algebra is worked out explicitly by using the 
diagrammatic notation in Appendix \ref{poss}.
The result is that the constraints form a set of first-class constraints, 
satisfying the following first-class Poisson algebra,
\[
&\{H (T_1),H  (T_2)\}
=J([\tilde T_1,\tilde T_2]+ 2\lambda\, T_1 \wedge T_2), \label{eq:HHJ}\\
&\{H(T),J(R) \} =H(T R), \label{eq:HJH} \\
&\{J(R_1),J(R_2) \} = J([R_1,R_2]), \label{eq:JJJ}
\]
where the matrix commutator is defined in \eq{eq:matrixcommutator}, 
$\tilde T_{ab} =T^cP_{cab}$, $(T_1 \wedge T_2)^{ab}
=T_1^a T_2^b-T_2^a T_1^b$,  and $(TR)_a=T^bR_{ba}$.

The consistency that the arguments of $J$ be anti-symmetric 
on the right-hand side of \eq{eq:HHJ} and \eq{eq:JJJ} can be checked from 
\[
&(\tilde T_1 \tilde T_2)^{ab}= 
T_1^c P_{c}{}^{ad} T_2^e P_{ed}{}^{b}
=(-1)^{(d+b)(a+d)+bd+ad+d}T_2^e P_{e}{}^{bd} T_1^c P_{cd}{}^{a} 
=(-1)^{ab} (\tilde T_2 \tilde T_1)^{ba}, \\
&(R_1  R_2)^{ab}= 
R_1^{ac} R_2{}_c{}^b=(-1)^{(a+c)(b+c)+ac+bc+c} R_2^{bc} R_1{}_c{}^a
=(-1)^{ab} (R_2 R_1)^{ba}. 
\]

In the case of $\lambda=0$, one can add another constraint given by
\[
D=\frac{1}{6} P^{abc}M_{cba}.
\label{eq:dilatation}
\]
This satisfies $\{D,P_{abc}\}=P_{abc},\ \{D,M_{abc}\}=-M_{abc}$, and 
$D$ is therefore a dilatational constraint. Such a dilatational constraint 
played a vital role in the connection between CTM and statistical systems 
on random networks, i.e.~randomly connected tensor networks 
\cite{Sasakura:2015xxa,Sasakura:2014zwa,Sasakura:2014yoa}.
From the dilatational property of $D$, it is obvious that $D$ forms 
a first-class constraint algebra with $H_0$ and $J$ as 
\[
&\{ D,H_0(T)\} =H_0(T), 
\label{eq:DH}\\
&\{D,J(R) \}=0. \label{eq:DJ}  
\]

\section{Reality condition}
\label{sec:reality}
%
The minimal set of dynamical variables of the purely bosonic CTM consist of a 
canonical conjugate pair of rank-three real symmetric tensors $M,P$. 
To consider this minimal setting also in the super-extended case, 
we will impose reality conditions on the super-extended variables in this section.
We will also check the consistency of the reality conditions with the constraints.

We denote the complex conjugation of a variable $A$ by $A^*$,
and impose that\footnote{We follow \cite{DeWitt:1992cy} for the convention.}
\[
(A_1 A_2 \ldots A_n)^*=A_n^* A_{n-1}^*\ldots A_1^*
\label{eq:conjtensor}
\]
for the complex conjugate of a product of variables $A_i$.
We consider $\Omega$ to be real for even index values, and pure imaginary 
for odd ones, respectively (note that, from the definition \eq{eq:omega}, 
$\Omega$ does not have Grassmann odd components.). From this demand, 
we obtain the complex conjugate of the $\Omega$:
\[
&\Omega_{ab}^*=(-1)^{ab} \Omega_{ab}=\Omega_{ba}, \CR
&\Omega^{ab}{}^*=(-1)^{ab} \Omega^{ab}=\Omega^{ba},
\label{eq:realityofOmega}
\]
where \eq{eq:omega} should also be reminded.
It is important to notice that the complex conjugation is generally not 
commutative with raising and lowering indices. This can be seen by
\[
(A_a)^*=(A^b \Omega_{ba})^*
=\Omega_{ab} A^{b*} \neq A^{b*} \Omega_{ba} \hbox{ in general}.
\label{eq:complexupper}
\]
Therefore, the definition of the complex conjugation of $A$ depends on whether
it is defined by upper or lower components.
We also obtain
\[
(A^a B_a)^*=(A_a \Omega^{ab} B_b)^*=B_b^* \Omega^{ba} A_a^*=B^{*a}A^*_a.
\label{eq:complexreverse}
\]
Graphically, this is 
\[
(A \middlerightarrow B)^*=B^* \middlerightarrow A^*,
\label{eq:reflection}
\]
namely, the complex conjugation is a reflection in the horizontal direction, but with 
reversed connections of arrows. Here, as mentioned above, it is important to define
the complex conjugations of variables for fixed positions (upper or lower) of their indices. 
Thus, throughout this paper, we define the complex conjugation of tensors (vectors and matrices as well) 
by its lower components. For example, $A^*{}_a=A_a^*$, and $A^{*a}=A_b^*\Omega^{ba}$, 
as used in \eq{eq:complexreverse}. One could define the complex conjugation of a 
variable by its upper components as well, but mixtures of the two ways of 
definitions would lead to unnecessary complications.

From the reflection property \eq{eq:reflection} of the complex conjugation, 
it is natural to define the reality/imaginary condition in the case of tensors 
to be the one dictating its relation with the tensor with reversed indices:
\[
T_{a_1 a_2 \ldots a_p}^*=\pm T_{a_p a_{p-1} \ldots a_1}.
\label{eq:reality}
\]
This is indeed what was imposed on $\Omega$ in \eq{eq:realityofOmega} 
with a plus sign. We impose the reality conditions on the dynamical variables as
\[
&M_{abc}^*=M_{cba},\CR
&P_{abc}^*=P_{cba}.
\label{eq:realityMP}
\]
The reality conditions on the non-dynamical variables are given by 
\[
&T_a^*=T_a,  \CR
&R_{ab}^*=-R_{ba}=(-1)^{ab} R_{ab},
\label{eq:realitynondynamical}
\]
where \eq{eq:defofanti} should also be reminded. Because of the anti-symmetry 
\eq{eq:defofanti} of $R$, we have to consider the reality condition with a 
minus sign in \eq{eq:realitynondynamical} to recover the usual reality 
condition in the purely bosonic case.

\begin{figure}
\begin{center}
\includegraphics[scale=.5]{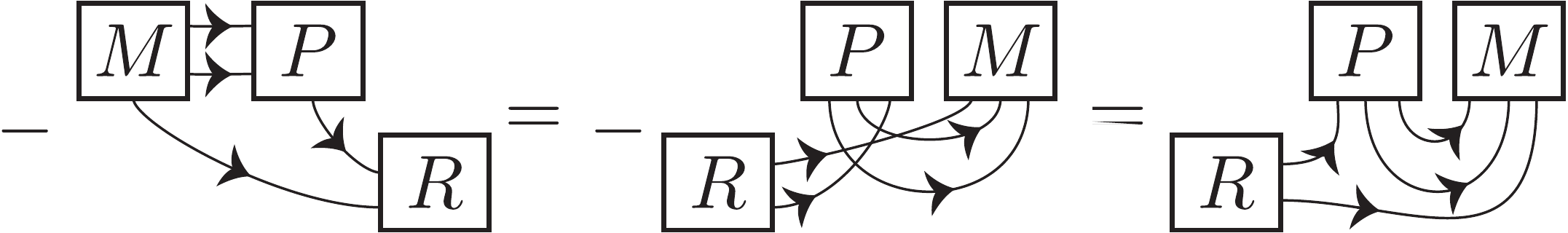}
\caption{The proof of reality of $J(R)$.
\label{fig:conjugateJ}}
\end{center}
\end{figure}
\begin{figure}
\begin{center}
\includegraphics[scale=.4]{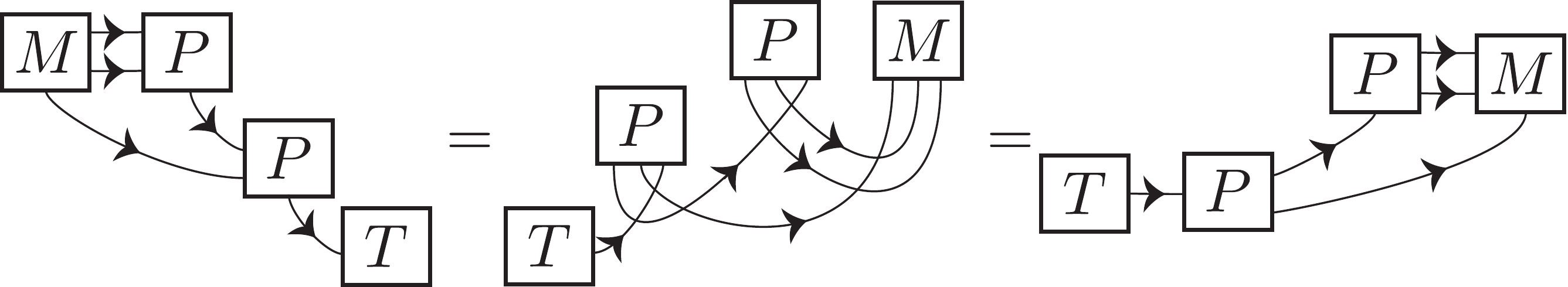}
\caption{The proof of reality of $H_0(T)$.
\label{fig:conjugateH}}
\end{center}
\end{figure}

The proof of the reality of the constraints is straightforward.
The proof for $J(R)$ is shown in Figure~\ref{fig:conjugateJ}. 
The first graph is obtained from the one of $J(R)$ in Figure~\ref{fig:hamiltonian} 
by taking its reflection with reversed arrows,
and its minus sign comes from \eq{eq:realitynondynamical}. 
The second one is obtained by using variants of the reordering 
identity Figure~\ref{fig:exchangerule}. Finally, the third one is obtained 
by the symmetric/anti-symmetric properties of the variables, 
Figure~\ref{fig:symcond} and \ref{fig:anti}. The proof for the reality of 
$H_0(T)$ proceeds basically along the same lines, as shown in 
Figure~\ref{fig:conjugateH}.
It is also easy to show the reality of the cosmological constant 
term in \eq{eq:hamiltonian}, and the dilatational constraint \eq{eq:dilatation}.

Finally, we should check whether the reality conditions are satisfied for 
the arguments of the constraints on the right-hand sides of the constraint 
algebra \eq{eq:HHJ}, \eq{eq:HJH}, and \eq{eq:JJJ}:
\[
&(\tilde T_1\tilde T_2)_{ab}^*=(T_1^c P_{ca}{}^d T_2^e P_{edb})^*=P_b{}^{de} T_{2e} P_{da}{}^c T_{1c}
=(-1)^{ab}T_1^c P_{ca}{}^d T_2^e P_{edb}=(-1)^{ab} (\tilde T_1 \tilde T_2)_{ab}, \CR
&(T_{1a} T_{2b})^*=T_{2b} T_{1a}=(-1)^{ab} T_{1a} T_{2b},\CR
&(TR)^*_a=(T^b R_{ba})^*=-R_{a}{}^b T_b=T^b R_{ba}, \CR
&(R_1R_2)_{ab}^*=(R_{1a}{}^c R_{2cb})^*=R_{2b}{}^cR_{1ca}=(-1)^{ab} R_{1a}{}^c R_{2cb}=(R_1R_2)_{ab},
\] 
where we have used \eq{eq:realityMP}, \eq{eq:realitynondynamical}, 
and \eq{eq:complexreverse} concerning the complex conjugation, 
and have reordered the variables and indices with attention to the 
signs using \eq{eq:orderfg}, \eq{eq:defofsym}, \eq{eq:defofanti}, and \eq{eq:contractorder}. 

\section{Quantization}
\label{sec:quantization}
%
In this section, we carry out the quantization of the classical system 
discussed so far. The quantized operator corresponding to a classical 
quantity $A$ will be denoted by $\hA$, and the same $Z_2$ grade 
will be assigned, i.e. $|\hat A|=|A|$. We assume the same symmetric 
properties \eq{eq:defofsym} for the quantized dynamical variables 
$\hM,\hP$. As for the reality condition, we impose 
\[
&\hM_{abc}^\dagger=\hM_{cba},\CR
&\hP_{abc}^\dagger=\hP_{cba},
\label{eq:hermiteMP}
\]
similar to \eq{eq:realityMP}, where $\hat A^\dagger$ denotes the 
Hermitian conjugate of $\hat A$. A reflection property similar to 
\eq{eq:complexreverse} holds, since
\[
(\hat f^a \hat g_a)^\dagger=(\hat f_a \Omega^{ab} \hat g_b)^\dagger
=\hat g_b^\dagger (\Omega^{ab})^* \hat f_a^\dagger
=\hat g_b^\dagger \Omega^{ba} \hat f_a^\dagger=
\hat g^{\dagger a} \hat f^\dagger_a,
\label{eq:hermitereverse}
\]
where we should note that $(\hat g^a){}^\dagger \neq \hat g^{\dagger a}$ 
in general, if $\hat g^\dagger$ is defined through its lower components. 
The definition of the Hermitian conjugate with lower components is assumed 
throughout this paper, as the complex conjugation in Section~\ref{sec:reality}.
 
The fundamental commutator is defined by
\[
& [\hM_{abc},\hP^{def} ] = - [\hP_{abc},\hM^{def} ]
= i\, \delta_{abc}^{fed}, \label{eq:fundamentalbracket}\\
& [\hM_{abc},\hM^{def} ]=[\hP_{abc},\hP^{def} ]=0,
\]
where $[\ ,\ ]$ is the operator commutator,
\[
[\hA, \hB]=\hA \hB-(-1)^{|A||B|} \hB \hA.
\]
The commutator satisfies
\[
[\hA, \hB]=-(-1)^{|A||B|}[\hB ,\hA].
\label{eq:commutationcommutator}
\]
By assuming the associativity of products of operators, 
one can readily prove the Leibniz rule,
\[
{}[\hA \hB,\hC]=\hA [\hB,\hC]+(-1)^{|B||C|} [\hA,\hC]\hB, \CR
{}[\hA, \hB\hC]=[\hA,\hB]\hC+(-1)^{|A||B|} \hB [\hA,\hC],
\label{eq:LeibnizCommutator}
\]
and the Jacobi identity in the operator language,
\[
{}[[\hA,\hB],\hC]=[\hA,[\hB,\hC]]+(-1)^{|B||C|} [[\hA,\hC],\hB].
\]

Note that the Leibniz rules \eq{eq:LeibnizCommutator} have the same form as 
the classical ones \eq{eq:LeibnizPoisson}. This means that the computation 
of the commutation algebra of the quantized constraints proceeds exactly 
in the same manner as the classical case\footnote{Except of course for 
the pre-factor $i$ in \eq{eq:fundamentalbracket}.}, if the orders of $\hM, \hP$ 
are not changed during the computation, since the classical Poisson algebra 
\eq{eq:HHJ}, \eq{eq:HJH}, \eq{eq:JJJ}, \eq{eq:DH}, and \eq{eq:DJ} of 
the constraints can in principle be derived solely by using \eq{eq:fundamentalpoisson}
and \eq{eq:LeibnizPoisson}. 
As will be discussed in due course, this fact will ensure that the classical Poisson algebra 
can be translated without any essential modifications to a quantum 
commutation algebra. 

As the reality condition on the classical constraints, we impose Hermiticity on the 
quantized constraints. By replacing the variables in \eq{eq:hamiltonian}, 
\eq{eq:momentum} with the quantized ones and taking into account the possible 
normal ordering term, the Hermiticity condition determines the quantized 
Hamiltonian and momentum constraints as
\[
&\hH(T)=\frac{1}{2} T^a \left( \hP_a{}^{bc} \hP_c{}^{de} \hM_{edb} +i \lambda_H \hP_{a}{}^b{}_b
-\lambda \hat M_a{}^b{}_b
 \right),
\label{eq:qhamiltonian}
\\
&\hJ(R)=-\frac{1}{2} R^{ab} \hP_{b}{}^{cd} \hM_{dca},
\label{eq:qmomentum}
\\
&\hD=\frac{1}{6} \left( \hP^{abc} \hM_{cba}+i \lambda_D \right) ,
\label{eq:qdilatation} 
\]
where $\lambda_H,\ \lambda_D$ are real constants coming from the 
normal ordering, and will be determined below. As will be checked below, 
there are no normal ordering terms for $\hJ(R)$. 

Let us first check the Hermiticity of $\hJ(R)$. By taking the Hermitian 
conjugate, we obtain
\[
\hJ(R)^\dagger&=-\frac{1}{2} \hM^{acd}\hP_{dc}{}^bR_{ba} \CR
&=\hJ(R)-\frac{1}{2} [\hM^{acd},\hP_{dc}{}^b]R_{ba}.
\label{eq:hermiteofhJ}
\]
where we have used \eq{eq:realitynondynamical}, \eq{eq:hermiteMP}, 
and \eq{eq:hermitereverse}.  
From \eq{eq:fundamentalbracket}, it is obvious that the second term can only produce 
either $R_{ab} \Omega^{ba}$ or $(-1)^a R_{ab}\Omega^{ba}=R_{ab} \Omega^{ab}$,
which both vanish due to the symmetry/anti-symmetry properties of $R,\Omega$. 
Therefore, \eq{eq:qmomentum} is Hermite.

\begin{figure}
\begin{center}
\includegraphics[scale=.4]{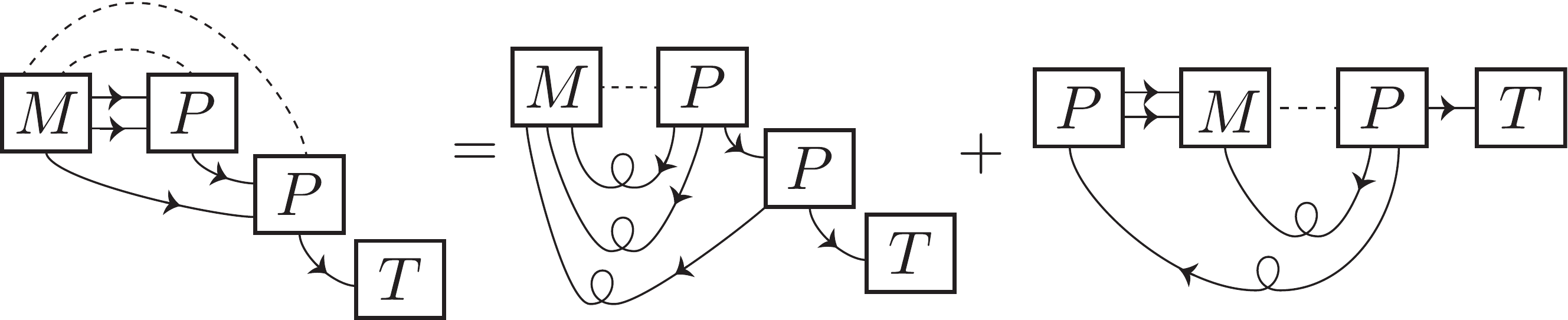}
\caption{Computation of the normal ordering term in $\hH(T)$. The dashed lines indicate the 
pairs of $\hM,\hP$, between which the fundamental commutators are taken. 
\label{fig:orderH}}
\end{center}
\end{figure}

\begin{figure}
\begin{center}
\includegraphics[scale=.7]{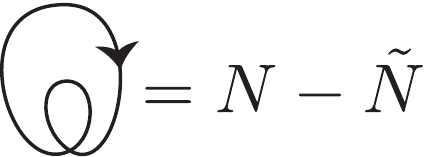}
\hfil
\includegraphics[scale=.7]{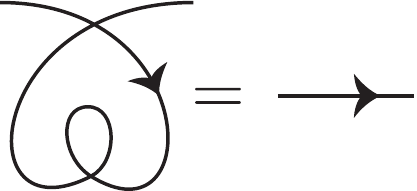}
\caption{
By substituting the fundamental commutator into Figure~\ref{fig:orderH}, one obtains 
some graphs containing the elementary ones shown in the figure.
\label{fig:Hnormal}}
\end{center}
\end{figure}

To study the Hermiticity of the quantized Hamiltonian constraint, let us take the 
Hermitian conjugate of the first term of \eq{eq:qhamiltonian}. Similarly as above, 
we obtain
\[
\left(T^a  \hP_a{}^{bc} \hP_c{}^{de} \hM_{edb} \right)^\dagger
=\hM^{bde} \hP_{ed}{}^c \hP_{cb}{}^a T_a 
=T^a  \hP_a{}^{bc} \hP_c{}^{de} \hM_{edb}+(\hbox{Figure~\ref{fig:orderH}}),
\]
where the normal ordering terms are shown in Figure~\ref{fig:orderH}.
Here, we have used some variants of the identity stated in Figure~\ref{fig:exchangerule},
and have also reversed the orientations of some of the arrows by using the rule 
given in Figure~\ref{fig:contractorder}. By making use of Figure~\ref{fig:poisson}, 
we obtain a sum of graphs which contain the elementary ones shown 
in Figure~\ref{fig:Hnormal}. The left graph produces a numerical factor 
$\sum_a (-1)^a =N-\tilde N$, while the right $(-1)^{2a} \delta_a^b=\delta_a^b$. 
By summing over all the contributions, we obtain 
\[
(\hbox{Figure}~\ref{fig:orderH})=(N-\tilde N +2)(N-\tilde N+3)\, T^a P_{a}{}^b{}_b.
\] 
Therefore, the Hermiticity of $\hH(T)$ is obtained by putting\footnote{The apparent difference by 6 of the numerical factors
of $\lambda_{H,D}$ from those in \cite{Narain:2014cya} come from the different normalization of the fundamental
commutator \eq{eq:fundamentalbracket}.} 
\[
\lambda_H=\frac{1}{2} (N-\tilde N +2)(N-\tilde N+3).
\]
One can see in this expression that the fermionic degrees of freedom contribute 
in an opposite manner to the bosonic ones, as commonly argued. 
In the same manner, we obtain
\[
\lambda_D=\frac{1}{2} (N-\tilde N)(N-\tilde N+1)(N-\tilde N+2).
\]

\begin{figure}
\begin{center}
\includegraphics[scale=.5]{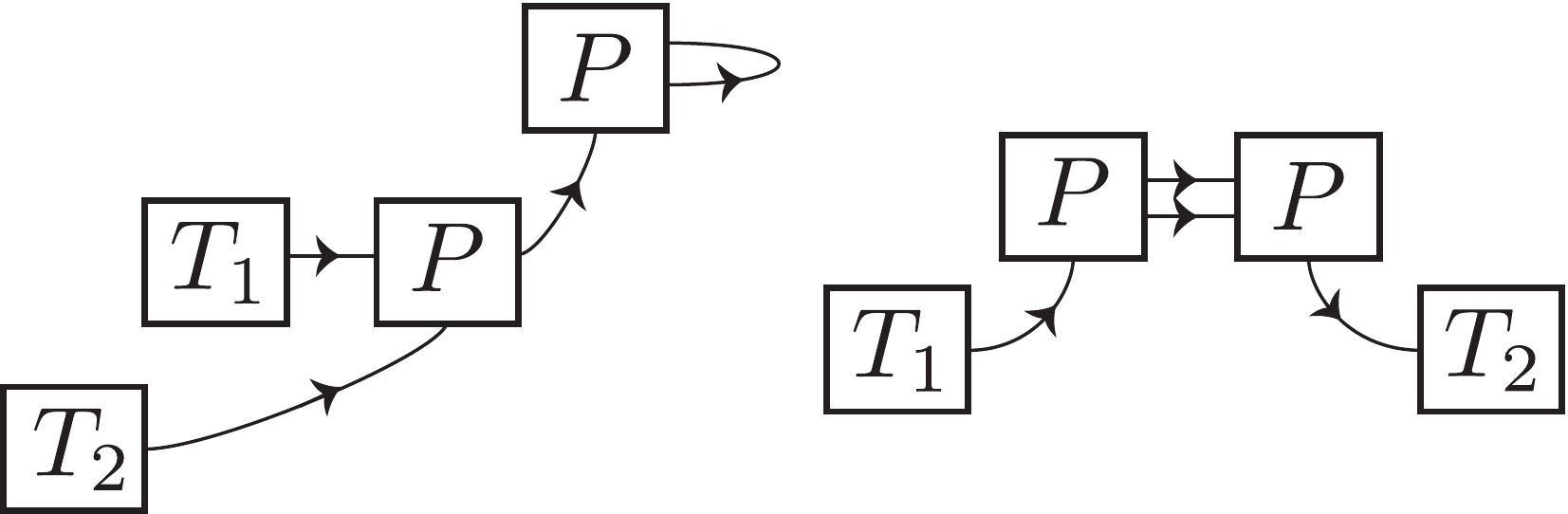}
\caption{The terms generated from the commutator 
$[T_1^a \hP_a{}^{bc} \hP_c{}^{de} \hM_{edb}, T_2^a   \hP_{a}{}^b{}_b]$.
\label{fig:normalextra}}
\end{center}
\end{figure}

As discussed above, the computation of the commutation algebra of the 
quantized constraints is basically the same as the classical one, if we keep 
the ordering of $\hM,\hP$, because of the formal equivalence between the 
identities satisfied by the Poisson bracket and the commutator. We take the 
normal ordering that $\hM$ be always on the rightmost if it exists, as the expressions of
$\hH(T),\hJ(R),\hD$ in \eq{eq:qhamiltonian}, \eq{eq:qmomentum}, \eq{eq:qdilatation}.
Under this convention of ordering, the same part of the quantized constraints as 
the classical ones generate the same algebra as the classical case. The only 
difference between the quantum and classical constraints is the existence of the normal 
ordering term proportional to $\hP_a{}^b{}_b$ in $\hH$ in \eq{eq:qhamiltonian}.
As we will see below, this extra term does not change the commutation algebra.
In Figure~\ref{fig:normalextra}, we show the graphs which are produced from 
the commutator with $\hH_0$, namely $[T_1^a \hP_a{}^{bc} \hP_c{}^{de} \hM_{edb}, T_2^a   \hP_{a}{}^b{}_b]$.
All the terms in Figure~\ref{fig:normalextra} are symmetric 
under $T_1\leftrightarrow T_2$, and therefore cancel out in the computation of $[\hH(T_1),\hH(T_2)]$.
Moreover, the normal ordering term does not either cause problems with 
the cosmological constant term, since the commutator between them can 
only produce the kinds of terms, $T_1^a T_{2a}$ or $T_1^a T_{2a}(-1)^a$, 
which are symmetric under $T_1 \leftrightarrow T_2$. One can also check 
the covariant property of the normal ordering term with $\hJ(R)$, and the 
commutator with $\hD$. Thus, we finally obtain the following quantum 
constraint commutation algebra, basically the same as the classical one, 
\[
&[\hH (T_1),\hH  (T_2)]=\hJ([\hat T_1,\hat T_2]+2 \lambda\, T_1 \wedge T_2), \label{eq:qHHJ} \\
&[\hH (T),\hJ(R) ] =\hH (T R), \label{eq:qHJH} \\
&[\hJ(R_1),\hJ(R_2) ]= \hJ([R_1,R_2]), \label{eq:qJJJ} \\
&[\hD,\hH_0(T)]=\hH_0(T), \\
&[\hD,\hJ(R)]=0,
\]
where $\hat T_{ab}= T^c \hP_{cab}$, and $\hH_0(T)=\hH(T)|_{\lambda=0}$. 
Here, note that the argument of $\hJ$ on the right-hand side of \eq{eq:qHHJ} contains $\hP$, 
and therefore is not commutative with the operator part of $\hat J$. 
Hence, we have to carefully put the argument on the leftmost as $R$ 
in the definition \eq{eq:qmomentum}.
As in the bosonic case, the commutation algebra is indeed first-class, and therefore
the quantization is consistent.

\section{Summary and discussions}
\label{sec:summary}

In this paper, we have made an attempt to include fermionic degrees of freedom
in CTM, which initially was purely bosonic in nature. We have introduced 
such degrees of freedom by allowing the dynamical rank-three tensors to 
be either Grassmann even or odd in accordance with the Grassmann natures associated to the indices.
This provides a straightforward super-extension of CTM, whose constraints and 
constraint algebra have basically the same form as the purely bosonic case 
except for the signs associated to the order of the indices and variables, 
extending the kinematical symmetry from $O(N)$ to $OSp(N,\tilde N)$.
To prove that the super-extended constraints form a first-class constraint algebra, we
have performed an explicit computations of the super-Poisson brackets.  
This process was facilitated by the graphical expressions which can 
represent the signs associated to the ordering in a simplified manner.
Then, we finally considered the quantization of the super-extended system. 
It has been shown that, as in the purely bosonic case, the quantized 
constraint algebra is formally the same as the classical one, which also means 
that the quantization is consistent. The quantized Hamiltonian constraint 
contains the normal ordering term proportional to $N-\tilde{N}$, which 
shows that the bosonic and fermionic degrees of freedom contribute oppositely. 

The formalism presented in this paper obviously contains the issue of 
negative norm states in the quantized case\footnote{For example, more details of this general aspect 
is explained in \cite{Henneaux:1992ig}.}. This issue comes 
from the Grassmann odd variables. From \eq{eq:fundamentalbracket} and 
some basic definitions, one can see that the Grassmann odd variables can 
be recast into the pairs $\psi^i,\tilde \psi^i$ which are hermite, 
$\psi^i=\psi^{i\,\dagger}, \tilde \psi^i=\tilde \psi^{i\,\dagger}$, and satisfy the 
anti-commutation relations, $[\psi^i,\tilde \psi^j]_+=\delta^{ij},\ [\psi^i,\psi^j]_+
=[\tilde \psi^i,\tilde \psi^j]_+=0$. Then, one can consider recombinations, 
$\psi^i_\pm=\psi^i\pm \tilde \psi^i$, and obtain $[\psi^i_-, \psi^j_-]_+=-2 \delta^{ij}$. 
Since $\psi^i_-$ is hermite and satisfies the last anti-commutation
relation, there exist negative norm states in the Hilbert space.

In order to resolve the issue in the quantized case, some 
directions of pursuit can be considered. 
One would be to add constraints like $\psi^i-\tilde \psi^i=0$ 
which incorporate the fact that the fermionic degrees of freedom should be 
treated in a first-order Lagrangian. 
This would require a careful analysis of the full consistency with the original constraints. 
Another more physically attractive direction would be to pursue the possibility to obtain fermionic degrees of 
freedom from quantization of bosonic degrees of freedom without initially 
introducing Grassmann odd variables (see for instance \cite{Rempel:2015foa} 
for a recent discussion in such directions). It would be fascinating, if the degrees 
of freedom of the purely bosonic CTM turn out to generate fermionic ones after 
quantization. Since fermionic degrees of freedom would be expected to appear 
as fluctuations around bosonic backgrounds, this would require some analysis 
of the behavior of the physical wave functions \cite{Narain:2014cya} in the vicinity 
of their peaks. Due to the spin-statistics theorem, it would even be possible that 
the above two different routes for resolution would finally lead to the same result. 
 
Leaving aside the issue in the quantized case, we would like to stress that the 
super-extended model presented in this paper itself would have some theoretical interests 
of its own. One is that, because of its formal equivalence to the purely bosonic 
case, it would be straightforward to extend the results obtained so far for the 
purely bosonic case. This would include the exact physical wave functions 
\cite{Sasakura:2013wza,Narain:2014cya} and the connection with the dual 
statistical systems, i.e. the randomly connected tensor networks 
\cite{Sasakura:2015xxa,Sasakura:2014zwa,Sasakura:2014yoa}.
Especially, since the connection of the latter is based on the classical 
aspects of CTM, the physical interpretations of the extension would 
be more solid. Although the additional degrees of freedom would not 
be real fermions, the super-extensions would provide general hints 
about how the anti-commuting degrees of freedom would change 
the physical outcomes compared with the purely bosonic case.

\appendix
\label{appen}

\section{Poisson algebra of the constraints}
\label{poss}

In the following subsections, we will prove the Poisson algebra, \eq{eq:HHJ}, 
\eq{eq:HJH}, and \eq{eq:JJJ} by explicit computations using the graphs introduced in 
Section~\ref{sec:graph}. 

\subsection{Computation of $\{J,J\}$}
\label{subsec:JJ}

\begin{figure}
[h]
\begin{center}
\includegraphics[scale=.4]{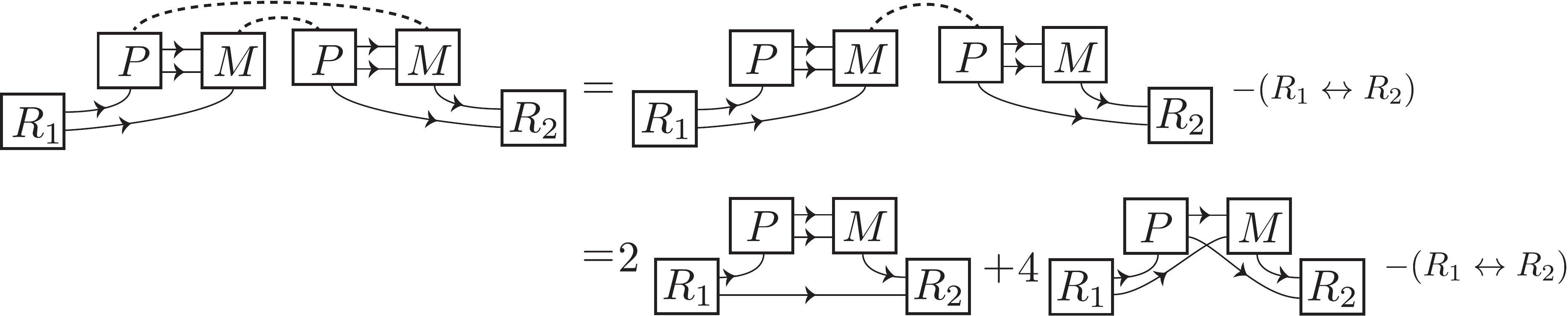}
\caption{The graphical computation of $\{J(R_1),J(R_2)\}$.
\label{fig:JJ}}
\end{center}
\end{figure}

\begin{figure}
\begin{center}
\includegraphics[scale=.4]{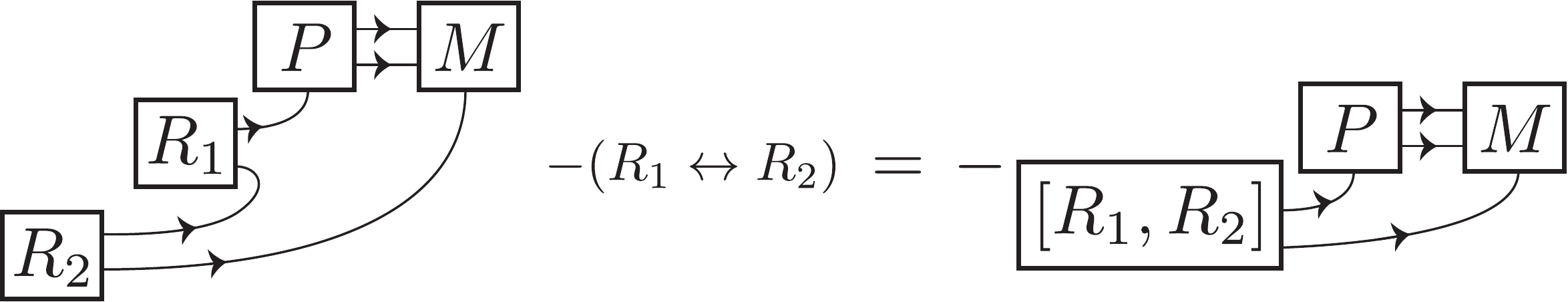}
\caption{The final result of the computation of Figure~\ref{fig:JJ}. 
The first graph can be obtained from the first one in the last line 
of Figure~\ref{fig:JJ} by using a variant of the identity Figure~\ref{fig:exchangerule}. 
\label{fig:finalJJ}}
\end{center}
\end{figure}

In this subsection, we will compute $\{ J(R_1),J(R_2) \}$. The computation of 
the Poisson bracket between $R_i^{ab}P_{b}{}^{cd}M_{dca}\ (i=1,2)$ is shown 
in Figure~\ref{fig:JJ}.
In the first graph, we have moved $R_2$ on the rightmost by using a 
variant of the identity Figure~\ref{fig:exchangerule} with two 
variables\footnote{Here, $P_{a}{}^{cd}M_{dcb}$ is regarded as one variable.}, and have 
indicated by the dashed lines the two cases of the fundamental Poisson bracket being taken.  
As shown in the second graph, one can easily find that the two cases contribute 
in opposite signs with $R_1\leftrightarrow R_2$. By substituting the fundamental 
Poisson bracket Figure~\ref{fig:poisson} into it and considering the symmetry 
Figure~\ref{fig:symcond} of $P,M$, we obtain the last line. In fact, the second 
graph of the last line is symmetric under $R_1 \leftrightarrow R_2$ and cancels out.
\begin{figure}
\begin{center}
\includegraphics[scale=.4]{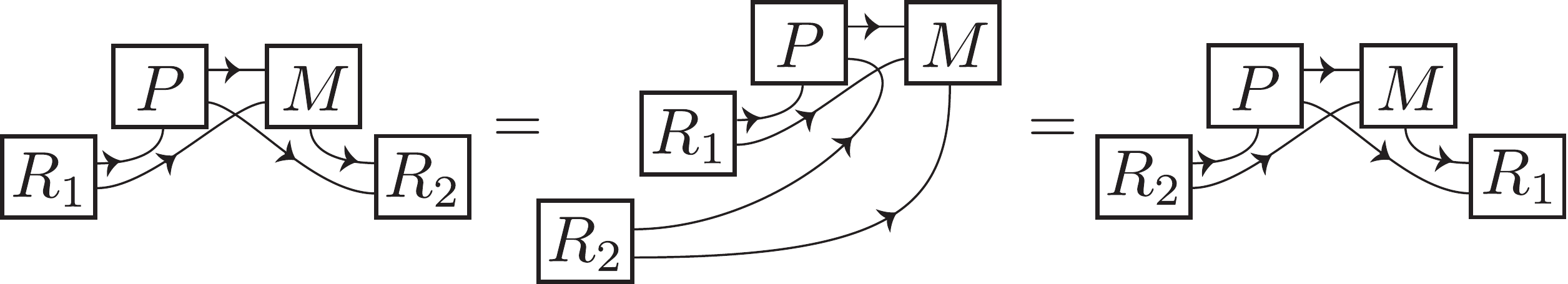}
\caption{By using a variant of the identity Figure~\ref{fig:exchangerule},
the second graph in the last line of Figure~\ref{fig:JJ}
can be shown to be symmetric under $R_1 \leftrightarrow R_2$.
\label{fig:changeR1R2}}
\end{center}
\end{figure}
The proof of the symmetric property of the second graph is shown in 
Figure~\ref{fig:changeR1R2}. We therefore finally obtain the result in 
Figure~\ref{fig:finalJJ}, where the first graph is obtained from the first 
one in the last line of Figure~\ref{fig:JJ} by moving $R_2$ on the leftmost 
with the use of a variant of the identity Figure~\ref{fig:exchangerule}. By taking 
care of the sign and the normalization in \eq{eq:momentum}, we obtain \eq{eq:JJJ}.

\subsection{Computation of $\{ H_0,H_0\}$}
%
In this subsection, we consider the $\lambda=0$ case first, and will deal with 
the $\lambda\neq 0$ case in Appendix~\ref{sec:cosmo}. 
\begin{figure}
[h]
\begin{center}
\includegraphics[scale=.4]{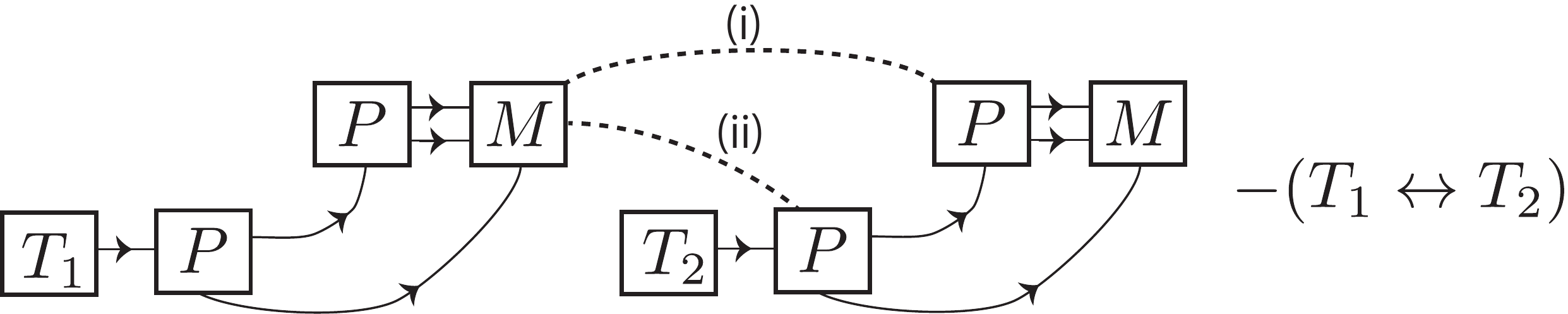}
\caption{The computation of $\{H_0,H_0\}$. There are two cases, (i) and (ii), of the fundamental
Poisson brackets to be taken.
\label{fig:HH}}
\end{center}
\end{figure}
As shown in Figure~\ref{fig:HH}, the Poisson bracket between 
$T_i^a P_a{}^{bc} P_c{}^{de}M_{edb} \ (i=1,2)$ is given by the summation 
of the contributions (i) and (ii) minus the same ones with $T_1 \leftrightarrow T_2$. 
By comparing with Figure~\ref{fig:JJ}, the computation of 
(i) minus $T_1 \leftrightarrow T_2$ of (i) is basically the same as the 
computation of $\{J,J\}$ with the replacement $(R_i){}_{ab}\rightarrow (\tilde T_i){}_{ab} =T_i^cP_{cab}$. 
Therefore, after taking care of the sign and the normalization, 
we obtain the right-hand side of \eq{eq:HHJ} with $\lambda=0$. 
Thus, what remains to show is that the contribution (ii) cancels out 
with $T_1 \leftrightarrow T_2$ of (ii). 
\begin{figure}
[h]
\begin{center}
\includegraphics[scale=.3]{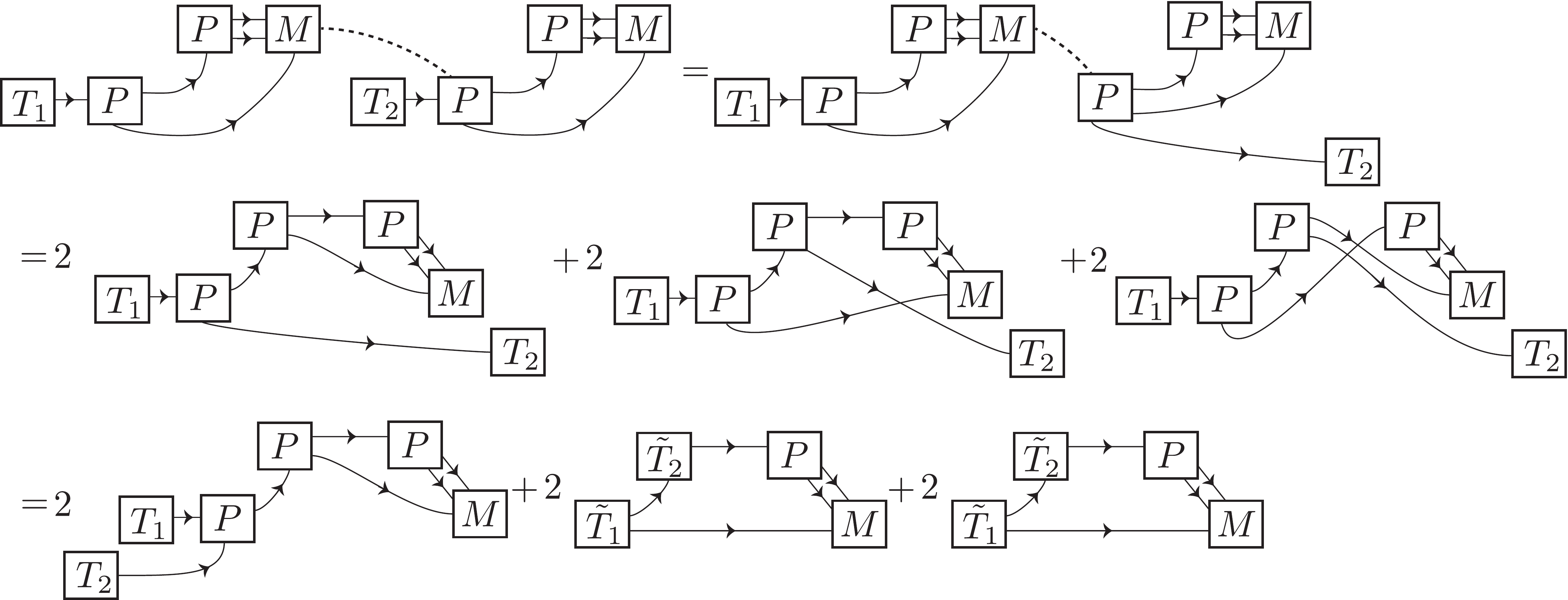}
\caption{The computation of the contribution (ii) in Figure~\ref{fig:HH}.
\label{fig:HHii}}
\end{center}
\end{figure}
\begin{figure}
[h]
\begin{center}
\includegraphics[scale=.3]{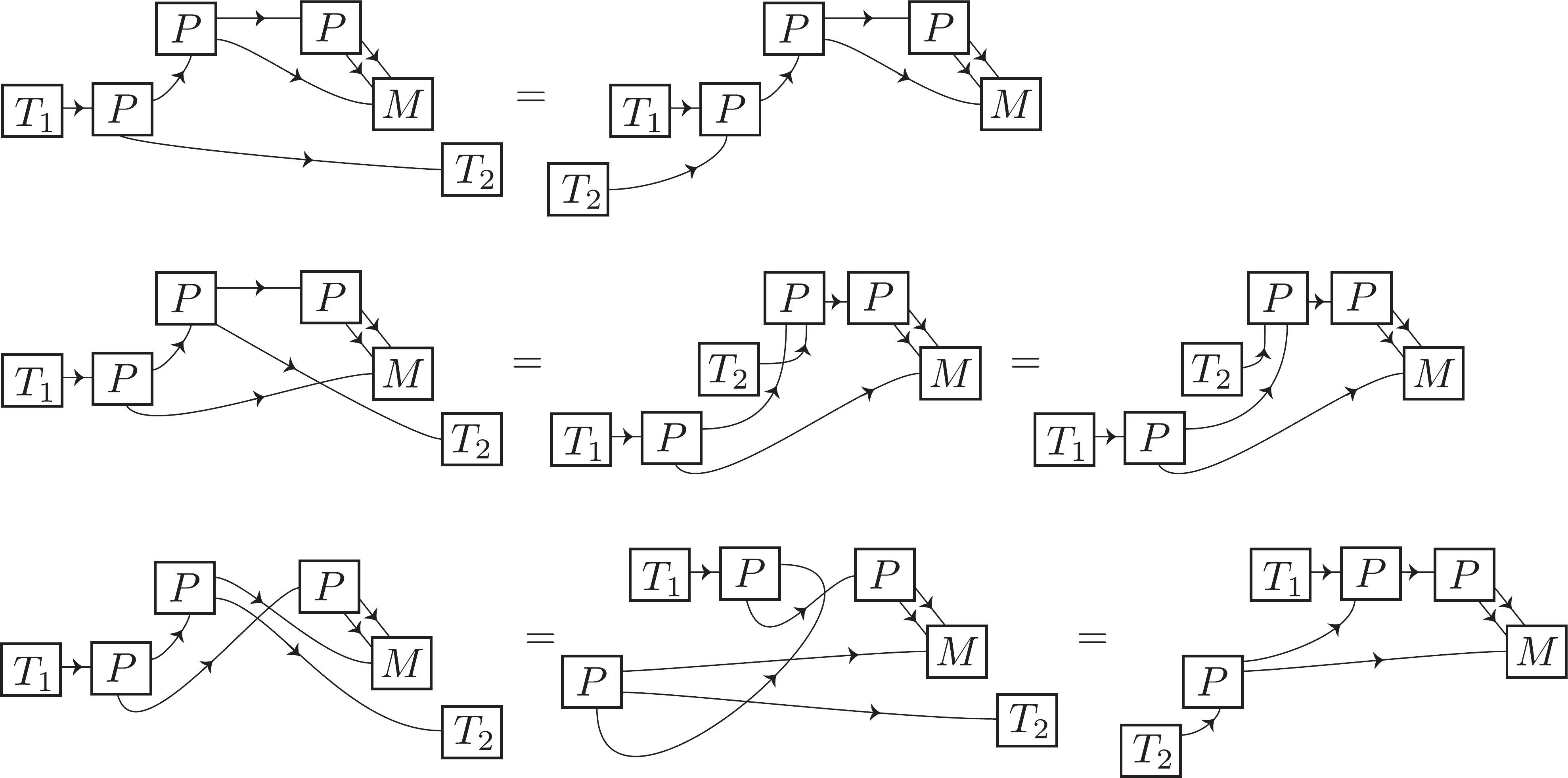}
\caption{From the second to the third line of Figure~\ref{fig:HHii}, each 
diagram is deformed by using variants of Figure~\ref{fig:exchangerule} 
and the symmetry Figure~\ref{fig:symcond} of $M,P$.
\label{fig:secondtothird}}
\end{center}
\end{figure}
In Figure~\ref{fig:HHii}, the computation of the contribution (ii) is shown.
In the first line, $T_2$ is moved on the rightmost by using a variant of the identity Figure~\ref{fig:exchangerule},
and then the fundamental Poisson bracket Fig.~\ref{fig:poisson} is substituted 
to obtain the second line. From the second to the last line, we use some variants 
of Figure~\ref{fig:exchangerule} and Figure~\ref{fig:symcond} to deform each diagram 
as shown in Figure~\ref{fig:secondtothird}. As shown in Figure~\ref{fig:T1T2exchange}, 
the first graph is invariant by itself under $T_1\leftrightarrow T_2$, and the sum of the 
last two graphs in Figure~\ref{fig:secondtothird} is obviously so, too. 
Therefore, the contribution (ii) cancels out with that of $T_1\leftrightarrow T_2$. 
\begin{figure}
[h]
\begin{center}
\includegraphics[scale=.4]{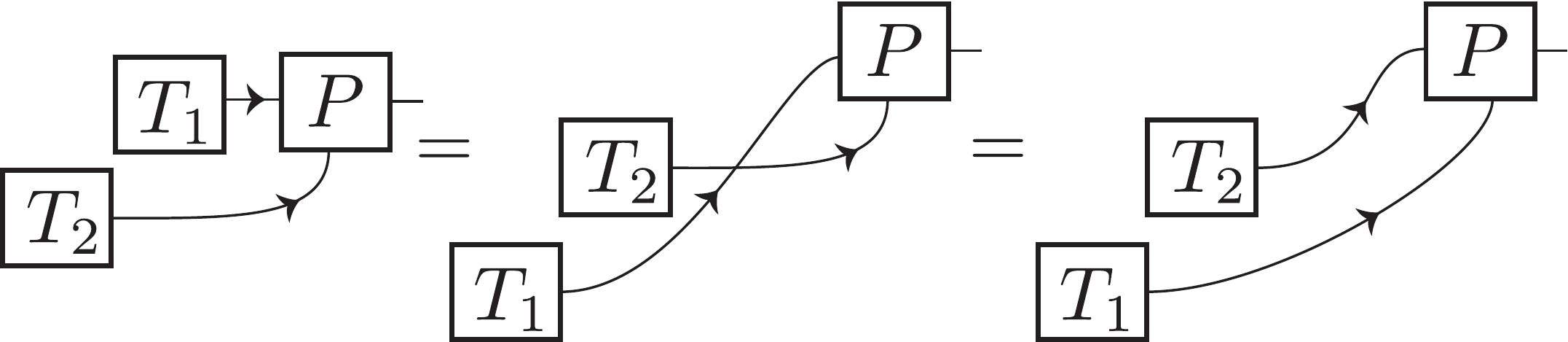}
\caption{The first graph of the last line in Figure~\ref{fig:HHii} is shown to be 
invariant under $T_1\leftrightarrow T_2$ by using a variant of Figure~\ref{fig:exchangerule} 
and Figure~\ref{fig:symcond}.
\label{fig:T1T2exchange}}
\end{center}
\end{figure}
%

\subsection{Computation of $\{H_0,J\}$}
\begin{figure}
[h]
\begin{center}
\includegraphics[scale=0.4]{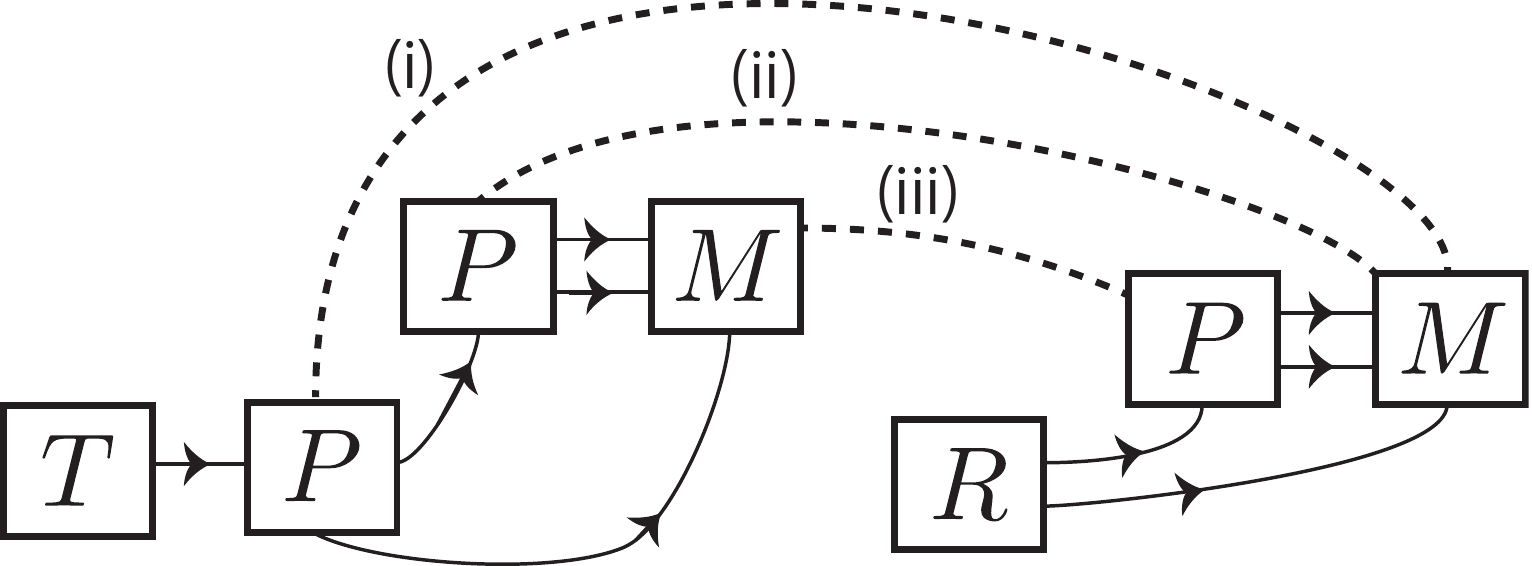}
\caption{For $\{ H_0(T),J(R)\}$, there exist three cases of the fundamental 
Poisson bracket being taken.
\label{fig:HJ}}
\end{center}
\end{figure}

%
In the computation of $\{H_0(T),J(R)\}$, there exist three cases of
the fundamental Poisson bracket to be taken, as shown in Figure~\ref{fig:HJ}.
\begin{figure}
\begin{center}
\includegraphics[height=6.6cm, width=12cm]{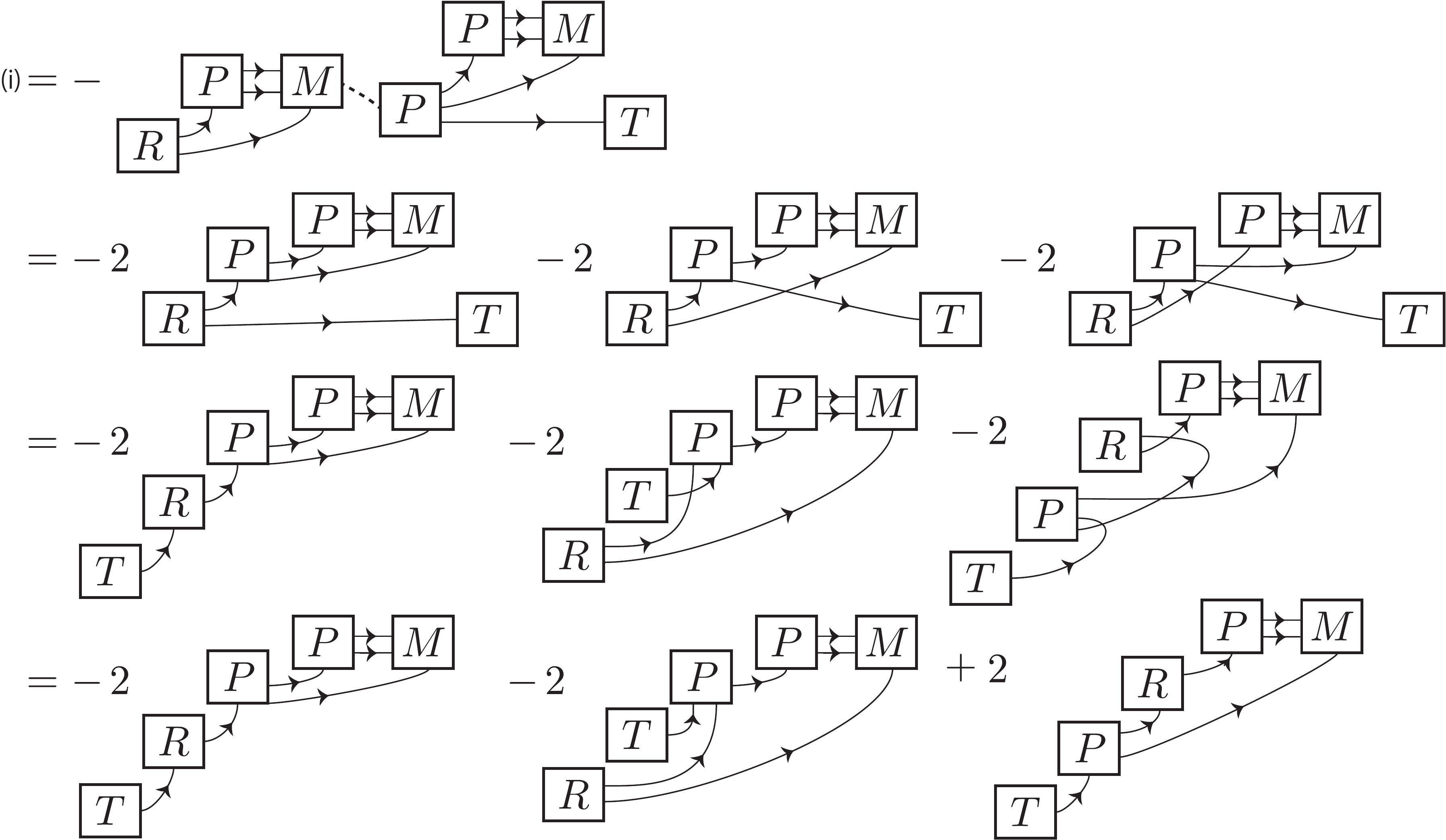}
\caption{The computation of the contribution (i) in Figure~\ref{fig:HJ}. In the first line, we perform 
an exchange of order, $\{J(R),H_0(T)\}=-\{H_0(T),J(R)\}$, and move $T$ on the rightmost 
by using a variant of Figure~\ref{fig:exchangerule}. In the second line,
the substitution of the fundamental Poisson bracket Figure~\ref{fig:poisson} results in a sum of three terms
due to the symmetric property of $P$. In the third line, some variants of 
Figure~\ref{fig:exchangerule} are used to move $T$ in the first and 
second terms and $T,P$ in the last term. Finally, by using the symmetric 
and anti-symmetric properties of $P$ and $R$, respectively, we obtain 
the last line. 
\label{fig:HJi}}
\end{center}
\end{figure}
The contributions (ii) and (iii) are the same as those appear in the computation of $\{J,J\}$ in
Figure~\ref{fig:JJ}, if we perform the replacement $R_1\rightarrow \tilde T,\  R_2\rightarrow R$.  
Thus the sum of (ii) and (iii) results in Figure~\ref{fig:finalJJ} with the replacement. 
On the other hand, the computation of the contribution (i) is shown in Figure~\ref{fig:HJi}.
By substituting the fundamental Poisson bracket and using some variants of the identity 
Figure~\ref{fig:exchangerule}, we obtain the expression of the last line of Figure~\ref{fig:HJi}.
Here the last two terms cancel with the contributions (ii) and (iii), namely, Figure~\ref{fig:finalJJ}
with $R_1\rightarrow \tilde T,\  R_2\rightarrow R$. Therefore, only the first term remains for 
$\{H_0(T),J(R)\}$. By taking care of the normalization and the sign of \eq{eq:momentum}, 
one obtains \eq{eq:HJH}.
%

\subsection{Cosmological constant term}
\label{sec:cosmo}
%
Here we consider the cosmological constant term. The difference 
between $\{ H,H\}$ and $\{ H_0,H_0\}$ is  
\[
\{H (T_1), H(T_2)\}
-\{H_0(T_1),H_0 (T_2)\}=
-\frac{\lambda}{2}\, \{H_0 (T_1),T_2^{a}  M_{a}{}^{bc}\Omega_{cb} \} 
- (T_1 \leftrightarrow T_2).
\]
\begin{figure}[h]
\begin{center}
\includegraphics[scale=.4]{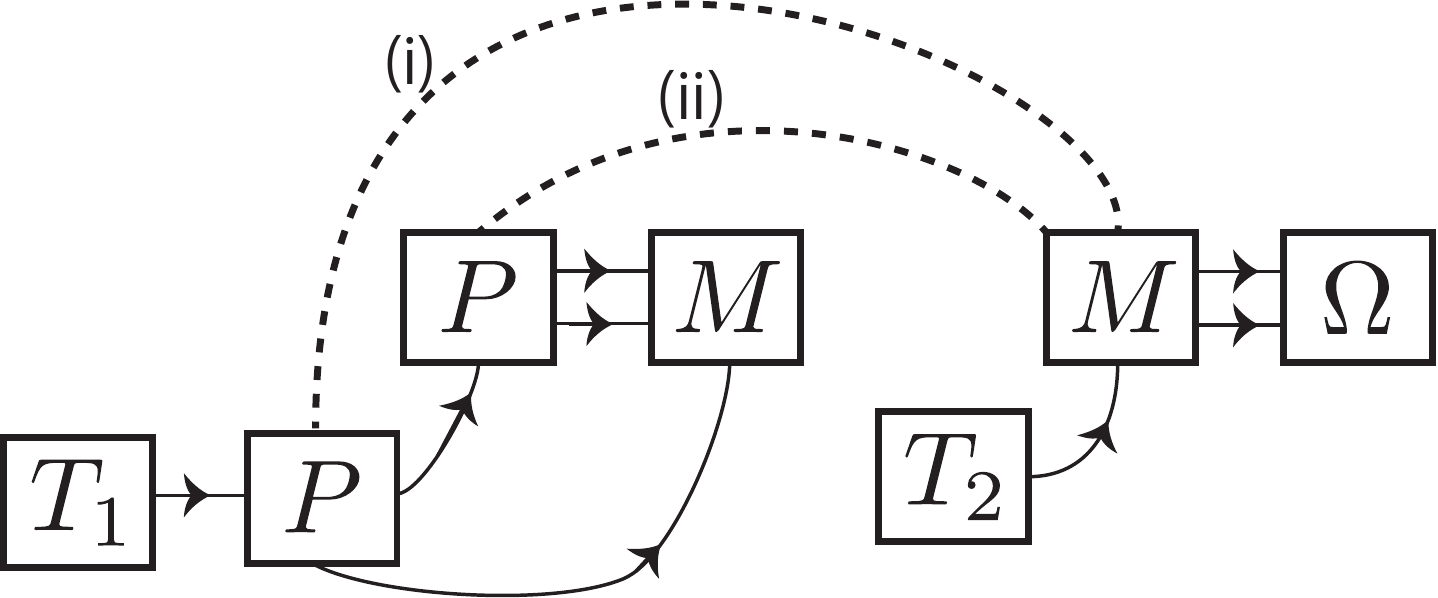}
\caption{The computation of $\{H_0 (T_1),T_2^{a}  M_{a}{}^{bc}\Omega_{cb} \}$. 
There are two contributions of the fundamental Poisson bracket. 
\label{fig:Hcosmo}}
\end{center}
\end{figure}
The computation of the first term on the right-hand side can be represented by 
Figure~\ref{fig:Hcosmo}. There are two contributions (i) and (ii) of the 
fundamental Poisson bracket. 
\begin{figure}
[!h]
\begin{center}
\includegraphics[scale=.4]{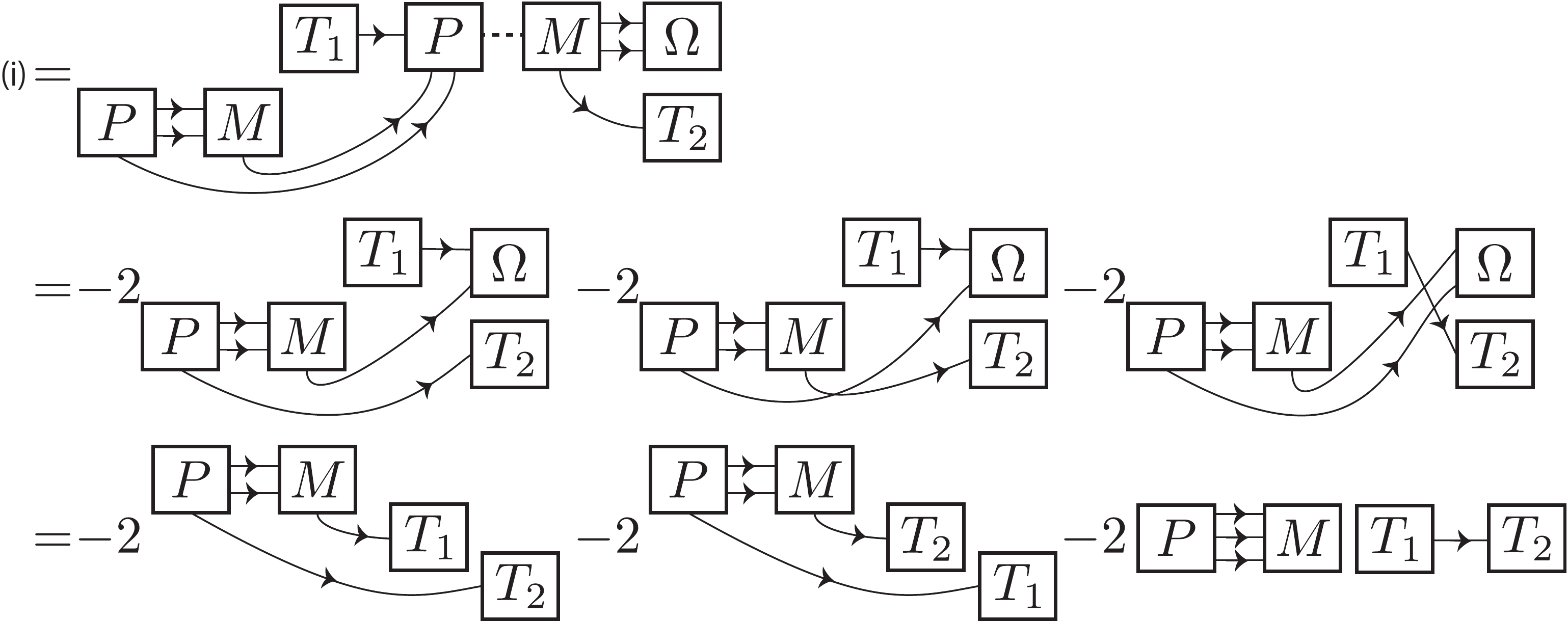}
\caption{The computation of (i) in Figure~\ref{fig:Hcosmo}. The result is symmetric 
under $T_1 \leftrightarrow T_2$, and cancels out.
\label{fig:Hcosmoi}}
\end{center}
\end{figure}
\begin{figure}
[!h]
\begin{center}
\includegraphics[scale=.4]{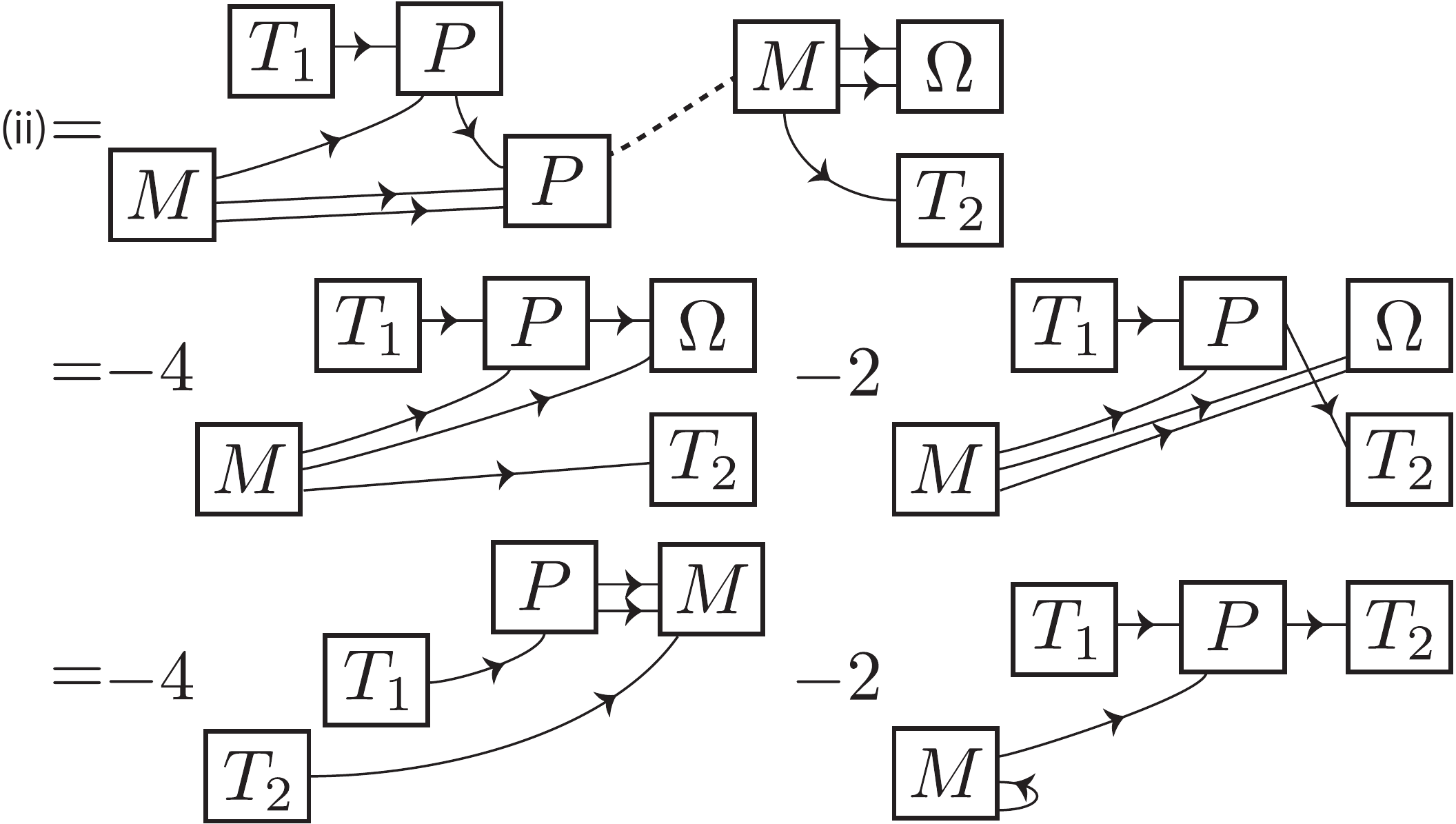}
\caption{The computation of (ii) in Figure~\ref{fig:Hcosmo}. The second term in the last line 
is symmetric under $T_1 \leftrightarrow T_2$, and cancels out. On the other hand, the 
first term generates the momentum constraint with the $T_1 \leftrightarrow T_2$ term.
\label{fig:Hcosmoii}}
\end{center}
\end{figure}
\begin{figure}
[!h]
\begin{center}
\includegraphics[scale=.4]{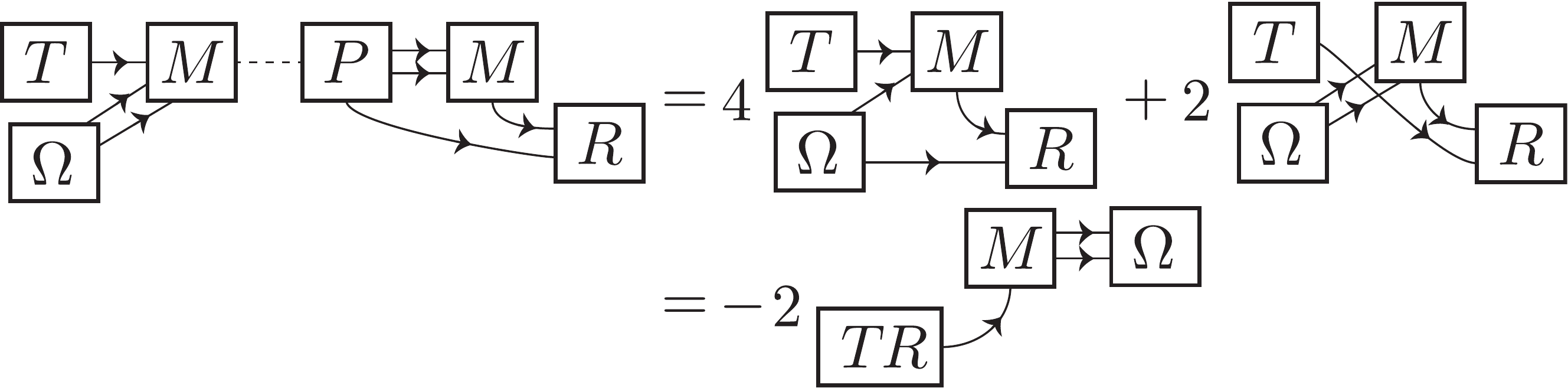}
\caption{The Poisson bracket between the cosmological constant term and $J(R)$. 
The first graph on the right-hand side of the first line vanishes because of the 
symmetry/anti-symmetry properties of $M,R$.
\label{fig:cosmoJ}}
\end{center}
\end{figure}
As shown by the graphical computation of Figure~\ref{fig:Hcosmoi},
the contribution (i) turns out to be symmetric under $T_1\leftrightarrow T_2$, 
and hence cancels out. Here, note that, since $\Omega$ is bosonic, we do 
not need to take care of the horizontal locations of $\Omega$ in the graphical expression.
The computation of the contribution (ii) is shown in Figure~\ref{fig:Hcosmoii}. 
The second term in the last line is symmetric and cancels out. On the other 
hand, with the $T_1\leftrightarrow T_2$ term, the first term generates 
$J(2 \lambda \, T_1 \wedge T_2)$
after taking into account the normalization and 
the sign. This is actually the term proportional to $\lambda$ on the right-hand side of \eq{eq:HHJ}.

We can also check \eq{eq:HJH}. The computation of the additional term proportional to $\lambda$ 
is given in Figure~\ref{fig:cosmoJ}.

\clearpage
\centerline{\bf Acknowledgements} 
The work of NS was supported in part by JSPS KAKENHI Grant Number 15K05050.
GN would like to thank NS for supporting his visit to YITP, where this work was started.
GN would like to thank his host Prof. K. S. Narain and Prof. Herman Nicolai for providing 
support and hospitality during his stay at ICTP and Max-Planck Institute for Gravitational Physics (AEI), 
Golm, respectively, where a part of the work was done.
NS would like to thank Daniele Oriti for the hospitality and support at his stay in the latter institute.
We are grateful to Yuki Sato for several useful discussions during the course of this work.


\begin{thebibliography}{40}

\bibitem{Ambjorn:1990ge}
  J.~Ambjorn, B.~Durhuus and T.~Jonsson,
  ``Three-Dimensional Simplicial Quantum Gravity And Generalized Matrix
  Models,''
  Mod.\ Phys.\ Lett.\ A {\bf 6}, 1133 (1991).

\bibitem{Sasakura:1990fs}
  N.~Sasakura,
  ``Tensor Model For Gravity And Orientability Of Manifold,''
  Mod.\ Phys.\ Lett.\ A {\bf 6}, 2613 (1991).

\bibitem{Godfrey:1990dt}
  N.~Godfrey and M.~Gross,
  ``Simplicial Quantum Gravity In More Than Two-Dimensions,''
  Phys.\ Rev.\ D {\bf 43}, 1749 (1991).
  
\bibitem{Fukuma:2015haa} 
  M.~Fukuma, S.~Sugishita and N.~Umeda,
  ``Putting matters on the triangle-hinge models,''
  arXiv:1504.03532 [hep-th].

\bibitem{Fukuma:2015xja} 
  M.~Fukuma, S.~Sugishita and N.~Umeda,
  ``Random volumes from matrices,''
  JHEP {\bf 1507}, 088 (2015)
  [arXiv:1503.08812 [hep-th]].
  
\bibitem{DiFrancesco:1993nw} 
  P.~Di Francesco, P.~H.~Ginsparg and J.~Zinn-Justin,
  ``2-D Gravity and random matrices,''
  Phys.\ Rept.\  {\bf 254}, 1 (1995)
  [hep-th/9306153].
  
\bibitem{Boulatov:1992vp}
  D.~V.~Boulatov,
  ``A Model of three-dimensional lattice gravity,''
  Mod.\ Phys.\ Lett.\ A {\bf 7}, 1629 (1992)
  [arXiv:hep-th/9202074].

\bibitem{Ooguri:1992eb}
  H.~Ooguri,
  ``Topological lattice models in four-dimensions,''
  Mod.\ Phys.\ Lett.\ A {\bf 7}, 2799 (1992)
  [arXiv:hep-th/9205090].
  
  \bibitem{DePietri:1999bx}
  R.~De Pietri, L.~Freidel, K.~Krasnov and C.~Rovelli,
  ``Barrett-Crane model from a Boulatov-Ooguri field theory over a  homogeneous
  space,''
  Nucl.\ Phys.\ B {\bf 574}, 785 (2000)
  [arXiv:hep-th/9907154].
  
\bibitem{Freidel:2005qe} 
  L.~Freidel,
  ``Group field theory: An Overview,''  Int.\ J.\ Theor.\ Phys.\  {\bf 44}, 1769 (2005)  [hep-th/0505016].  

\bibitem{Oriti:2011jm} 
  D.~Oriti,
  ``The microscopic dynamics of quantum space as a group field theory,''  arXiv:1110.5606 [hep-th].  

\bibitem{DePietri:2000ii} 
  R.~De Pietri and C.~Petronio,
  ``Feynman diagrams of generalized matrix models and the associated manifolds in dimension 4,''
  J.\ Math.\ Phys.\  {\bf 41}, 6671 (2000)
  [gr-qc/0004045].

\bibitem{Gurau:2009tw} 
  R.~Gurau,
  ``Colored Group Field Theory,''  Commun.\ Math.\ Phys.\  {\bf 304}, 69 (2011)  
  [arXiv:0907.2582 [hep-th]].  

\bibitem{Gurau:2011xp} 
  R.~Gurau and J.~P.~Ryan,
  ``Colored Tensor Models - a review,''
  SIGMA {\bf 8}, 020 (2012)
  [arXiv:1109.4812 [hep-th]].  
  
  
\bibitem{Benedetti:2015ara} 
  D.~Benedetti and R.~Gurau,
  ``Symmetry breaking in tensor models,''
  arXiv:1506.08542 [hep-th].
  
\bibitem{Delepouve:2015nia} 
  T.~Delepouve and R.~Gurau,
  ``Phase Transition in Tensor Models,''
  JHEP {\bf 1506}, 178 (2015)
  [arXiv:1504.05745 [hep-th]].
  
\bibitem{Bonzom:2015axa} 
  V.~Bonzom, T.~Delepouve and V.~Rivasseau,
  ``Enhancing non-melonic triangulations: A tensor model mixing melonic and planar maps,''
  Nucl.\ Phys.\ B {\bf 895}, 161 (2015)
  [arXiv:1502.01365 [math-ph]].
  
\bibitem{Delepouve:2014hfa} 
  T.~Delepouve and V.~Rivasseau,
  ``Constructive Tensor Field Theory: The $T^4_3$ Model,''
  arXiv:1412.5091 [math-ph].
  
\bibitem{Nguyen:2014mga} 
  V.~A.~Nguyen, S.~Dartois and B.~Eynard,
  ``An analysis of the intermediate field theory of T$^4$ tensor model,''
  JHEP {\bf 1501}, 013 (2015)
  [arXiv:1409.5751 [math-ph]].
    
 
\bibitem{Benedetti:2015yaa} 
  D.~Benedetti and V.~Lahoche,
  ``Functional Renormalization Group Approach for Tensorial Group Field Theory: A Rank-6 Model with Closure Constraint,''
  arXiv:1508.06384 [hep-th].
   
\bibitem{Avohou:2015sia} 
  R.~C.~Avohou, V.~Rivasseau and A.~Tanasa,
  ``Renormalization and Hopf Algebraic Structure of the 5-Dimensional Quartic Tensor Field Theory,''
  arXiv:1507.03548 [math-ph].
 
\bibitem{Geloun:2015lta} 
  J.~B.~Geloun,
  ``A power counting theorem for a $p^{2a}\phi^4$ tensorial group field theory,''
  arXiv:1507.00590 [hep-th].
 
\bibitem{Lahoche:2015ola} 
  V.~Lahoche, D.~Oriti and V.~Rivasseau,
  ``Renormalization of an Abelian Tensor Group Field Theory: Solution at Leading Order,''
  JHEP {\bf 1504}, 095 (2015)
  [arXiv:1501.02086 [hep-th]].
  
\bibitem{Benedetti:2014qsa} 
  D.~Benedetti, J.~Ben Geloun and D.~Oriti,
  ``Functional Renormalisation Group Approach for Tensorial Group Field Theory: a Rank-3 Model,''
  JHEP {\bf 1503}, 084 (2015)
  [arXiv:1411.3180 [hep-th]].
  
\bibitem{Geloun:2014ema} 
  J.~Ben Geloun and R.~Toriumi,
  ``Parametric Representation of Rank d Tensorial Group Field Theory: Abelian Models with Kinetic Term $\sum_{s}|p_s| + \mu$,''
  arXiv:1409.0398 [hep-th].
       
  
\bibitem{Bonzom:2011zz} 
  V.~Bonzom, R.~Gurau, A.~Riello and V.~Rivasseau,
  ``Critical behavior of colored tensor models in the large N limit,''
  Nucl.\ Phys.\ B {\bf 853}, 174 (2011)
  [arXiv:1105.3122 [hep-th]].
  
\bibitem{Gurau:2013cbh} 
  R.~Gurau and J.~P.~Ryan,
  ``Melons are branched polymers,''
  Annales Henri Poincare {\bf 15}, no. 11, 2085 (2014)
  [arXiv:1302.4386 [math-ph]].
  

\bibitem{Raasakka:2013eda} 
  M.~Raasakka and A.~Tanasa,
  ``Next-to-leading order in the large $N$ expansion of the multi-orientable random tensor model,''
  Annales Henri Poincare {\bf 16}, no. 5, 1267 (2015)
  [arXiv:1310.3132 [hep-th]].

\bibitem{Dartois:2013sra} 
  S.~Dartois, R.~Gurau and V.~Rivasseau,
  ``Double Scaling in Tensor Models with a Quartic Interaction,''
  JHEP {\bf 1309}, 088 (2013)
  [arXiv:1307.5281 [hep-th]].

\bibitem{Kaminski:2013maa} 
  W.~Kami\'nski, D.~Oriti and J.~P.~Ryan,
  ``Towards a double-scaling limit for tensor models: probing sub-dominant orders,''
  New J.\ Phys.\  {\bf 16}, 063048 (2014)
  [arXiv:1304.6934 [hep-th]].

\bibitem{Gurau:2013pca} 
  R.~Gurau,
  ``The 1/N Expansion of Tensor Models Beyond Perturbation Theory,''
  Commun.\ Math.\ Phys.\  {\bf 330}, 973 (2014)
  [arXiv:1304.2666 [math-ph]].
  
\bibitem{Ambjorn:2004qm}
  J.~Ambjorn, J.~Jurkiewicz and R.~Loll,
  ``Emergence of a 4-D world from causal quantum gravity,''
  Phys.\ Rev.\ Lett.\  {\bf 93} (2004) 131301
  [hep-th/0404156].
  
\bibitem{Sasakura:2011sq}
  N.~Sasakura,
  ``Canonical tensor models with local time,''
  Int.\ J.\ Mod.\ Phys.\ A {\bf 27} (2012) 1250020
  [arXiv:1111.2790 [hep-th]].

\bibitem{Sasakura:2012fb}
  N.~Sasakura,
  ``Uniqueness of canonical tensor model with local time,''
  Int.\ J.\ Mod.\ Phys.\ A {\bf 27} (2012) 1250096
  [arXiv:1203.0421 [hep-th]].

\bibitem{Sasakura:2013gxg}
  N.~Sasakura,
  ``A canonical rank-three tensor model with a scaling constraint,''
  Int.\ J.\ Mod.\ Phys.\ A {\bf 28} (2013) 1
  [arXiv:1302.1656 [hep-th]].
  
\bibitem{Oriti:2013aqa} 
  D.~Oriti,
  ``Group field theory as the 2nd quantization of Loop Quantum Gravity,''
  arXiv:1310.7786 [gr-qc].
  
\bibitem{Arnowitt:1960es}
  R.~L.~Arnowitt, S.~Deser and C.~W.~Misner,
  ``Canonical variables for general relativity,''
  Phys.\ Rev.\  {\bf 117}, 1595 (1960).
  
\bibitem{Arnowitt:1962hi}
  R.~L.~Arnowitt, S.~Deser and C.~W.~Misner,
  ``The Dynamics of general relativity,''
  arXiv:gr-qc/0405109.
  
\bibitem{DeWitt:1967yk} 
  B.~S.~DeWitt,
  ``Quantum Theory of Gravity. 1. The Canonical Theory,''
  Phys.\ Rev.\  {\bf 160}, 1113 (1967).

\bibitem{Hojman:1976vp} 
  S.~A.~Hojman, K.~Kuchar and C.~Teitelboim,
  ``Geometrodynamics Regained,''
  Annals Phys.\  {\bf 96}, 88 (1976).
  
\bibitem{Teitelboim:1987zz} 
  C.~Teitelboim and J.~Zanelli,
  ``Dimensionally continued topological gravitation theory in Hamiltonian form,''
  Class.\ Quant.\ Grav.\  {\bf 4}, L125 (1987).

\bibitem{Sasakura:2014gia} 
  N.~Sasakura and Y.~Sato,
  ``Interpreting canonical tensor model in minisuperspace,''
  Phys.\ Lett.\ B {\bf 732}, 32 (2014)
  [arXiv:1401.2062 [hep-th]].
  
\bibitem{Sasakura:2015pxa} 
  N.~Sasakura and Y.~Sato,
  ``Constraint algebra of general relativity from a formal continuum limit of canonical tensor model,''
  arXiv:1506.04872 [hep-th].
  

\bibitem{Sasakura:2015xxa} 
  N.~Sasakura and Y.~Sato,
  ``Renormalization procedure for random tensor networks and the canonical tensor model,''
  PTEP {\bf 2015}, no. 4, 043B09 (2015)
  [arXiv:1501.05078 [hep-th]].
  
\bibitem{Sasakura:2014zwa} 
 N.~Sasakura and Y.~Sato,
  ``Ising model on random networks and the canonical tensor model,''
  PTEP {\bf 2014}, no. 5, 053B03 (2014)
  [arXiv:1401.7806 [hep-th]].
  
\bibitem{Sasakura:2014yoa} 
  N.~Sasakura and Y.~Sato,
  ``Exact Free Energies of Statistical Systems on Random Networks,''
  SIGMA {\bf 10}, 087 (2014)
  [arXiv:1402.0740 [hep-th]].
  
\bibitem{Sasakura:2013wza}
  N.~Sasakura,
  ``Quantum canonical tensor model and an exact wave function,''
  Int.\ J.\ Mod.\ Phys.\ A {\bf 28} (2013) 1350111
  [arXiv:1305.6389 [hep-th]].
  
\bibitem{Narain:2014cya} 
  G.~Narain, N.~Sasakura and Y.~Sato,
  ``Physical states in the canonical tensor model from the perspective of random tensor networks,''
  JHEP {\bf 1501}, 010 (2015)
  [arXiv:1410.2683 [hep-th]].
  
\bibitem{Sasakura:2008pe} 
  N.~Sasakura,
  ``Emergent general relativity on fuzzy spaces from tensor models,''
  Prog.\ Theor.\ Phys.\  {\bf 119}, 1029 (2008)
  [arXiv:0803.1717 [gr-qc]].

\bibitem{Sasakura:2009hs} 
  N.~Sasakura,
  ``Gauge fixing in the tensor model and emergence of local gauge symmetries,''
  Prog.\ Theor.\ Phys.\  {\bf 122}, 309 (2009)
  [arXiv:0904.0046 [hep-th]].
  
\bibitem{Rempel:2015foa} 
  T.~Rempel and L.~Freidel,
  ``Interaction Vertex for Classical Spinning Particles,''
  arXiv:1507.05826 [hep-th].
  
  
\bibitem{DeWitt:1992cy} 
  B.~S.~DeWitt,
  ``Supermanifolds,''
  Cambridge Monographs on Mathematical Physics, Cambridge University Press (1992) 407 p.
  
\bibitem{Henneaux:1992ig} 
  M.~Henneaux and C.~Teitelboim,
  ``Quantization of gauge systems,''
  Princeton, USA: Univ. Pr. (1992) 520 p.
  
\end{thebibliography}
\end{document}